\documentclass[11pt, a4paper]{article}
\usepackage[margin=2cm]{geometry}
\usepackage{times,graphicx,amsmath,amssymb,natbib,authblk,lineno,setspace,mathrsfs}
\usepackage[lofdepth,lotdepth]{subfig}
\usepackage{hyperref}

\title{The melting column as a filter of mantle trace-element
  heterogeneity} 

\author{Tong Bo$^0$, Richard F.~Katz$^1$, Oliver
  Shorttle$^{2,3}$ \& John F. Rudge$^{4}$}

\affil{$^0$ Peking University, Beijing, China (Current address:
  MIT-WHOI Joint Program, Cambridge, MA 02139, USA); $^1$ Department
  of Earth Sciences, University of Oxford, South Parks Road, Oxford
  OX1 3AN, United Kingdom; $^2$ Department of Earth Sciences,
  University of Cambridge, Downing Street, Cambridge CB2 3EQ, United
  Kingdom; $^3$ Institute of Astronomy, University of Cambridge,
  Madingley Road, Cambridge CB3 0HA, United Kingdom. $^4$ Bullard
  Laboratories, Department of Earth Sciences, University of Cambridge,
  Madingley Road, Cambridge CB3 0EZ, United Kingdom.}  \affil{email:
  richard.katz@earth.ox.ac.uk}

\newcommand{\infd}{\text{d}}
\newcommand{\e}{\text{e}}
\newcommand{\diff}[2]{\frac{\infd{#1}}{\infd{#2}}}
\newcommand{\pdiff}[2]{\frac{\partial{#1}}{\partial{#2}}}
\newcommand{\Ldiff}[3]{\frac{\text{D}_{#1}#2}{\text{D}#3}}
\newcommand{\cs}{c^s}
\newcommand{\cl}{c^\ell}
\newcommand{\csm}{\overline{c}^s}
\newcommand{\clm}{\overline{c}^\ell}
\newcommand{\csf}{\breve{c}^s}
\newcommand{\clf}{\breve{c}^\ell}
\newcommand{\vcon}{C}
\newcommand{\Tsol}{T^S}
\newcommand{\Fmax}{F_\text{max}}
\newcommand{\diffusivity}{\mathcal{D}^s}
\newcommand{\clapeyron}{\gamma}
\newcommand{\cg}{c^\Gamma}
\newcommand{\rxn}{\mathcal{X}}   
\newcommand{\pco}{K}             
\newcommand{\rr}{\mathcal{R}}    
\newcommand{\admit}{\mathcal{A}} 
\newcommand{\admitl}{\admit^\ell} 
\newcommand{\admits}{\admit^s}
\newcommand{\iprod}{\Pi}         
\newcommand{\freq}{\Omega}
\newcommand{\real}{\text{Re}}

\newcommand{\dura}{\mathcal{G}}
\newcommand{\cvar}{\mathcal{C}_{\textrm{var}}}
\newcommand{\sha}{p}

\begin{document}
\maketitle
\doublespace

\begin{abstract}
  The observed variability of trace-element concentration in basaltic
  lavas and melt inclusions carries information about heterogeneity in
  the mantle. The difficulty is to disentangle the contributions of
  source heterogeneity (i.e., spatial variability of mantle
  composition before melting) and process heterogeneity (i.e., spatial
  and temporal variability in melt transport).  Here we investigate
  the end-member hypothesis that variability arises due to source
  heterogeneity alone.  We model the attenuation of trace-element
  variability introduced into the bottom of a one-dimensional,
  steady-state melting column. Our results show that the melting
  column can be considered to be a filter that attenuates variability
  according to the wavelength of heterogeneity, the partition
  coefficient of the trace element, melt productivity and the
  efficiency of melt segregation. We further show that while the model
  can be fit to the observations, this requires assumptions inconsistent
  with constraints on the timescales of magma assembly. Hence, we falsify
  the end-member hypothesis and,  instead, conclude that observed
  variability requires heterogeneity of melt transport.  This might
  take the form of channels or waves and would almost certainly
  interact with source heterogeneity.
\end{abstract}

\section{Introduction}

Basaltic lava compositions can potentially constrain models of
melting, melt transport, and the chemical character of the source
mantle.  Increasing attention has focused on the meaning of chemical
variability at different length scales \citep[e.g,][]{gurnis86,
  allegre86, laubier12, shorttle15, neave18}.  Some of this
variability is inherited from mantle-derived magmas that are the
product of melting a heterogeneous source. The melting process and
melt transport determine how that source is sampled by the segregating
magma. Therefore, observed geochemical variability should contain a
signal that represents a (conceptual) convolution of source and
process.  Deconvolving these requires a quantitative understanding of
the factors that determine magma composition.

Two end-members of such models could be imagined. In the first,
the mantle source, prior to any melting, is homogeneous. Physical
instability leads to spatio-temporal variability of melt
transport. The most prominent example is of channelised magmatic flow,
arising by a reactive instability \citep[e.g.,][]{kelemen95,
  aharonov95, spiegelman01, liang10b, reesjones18b}. Channels are
hypothesised to transport deep, low-degree, enriched melts to the
surface without aggregating the depleted melts that are produced at
shallower depths \citep{spiegelman03a}. Magmatic solitary waves may
also be capable of transporting deep, enriched melts in isolation from
those produced shallower \citep{jordan18}.

The other end-member is a heterogeneous mantle source with uniform
melt transport.  In this case, it is sufficient to consider a model
domain that is one-dimensional, aligned with the vertical. This
end-member can address only trace-element or isotope heterogeneity, as
these do not modify the melting rate (as would major-element
heterogeneity). Lithophile trace elements have low concentrations in
the source, typically slow diffusivities in solid phases and distinct
incompatibilities; they provide a useful indicator of magmatic
processes.  Large variations in the concentration of incompatible
lithophile trace elements are routinely observed in suites of primitive basalts
and melt inclusions. Isotopic evidence requires that some of this
variation must be inherited from the mantle source
\citep[e.g.,][]{saal1998_science, stracke03, maclennan2008_gca}.  The
model developed below addresses this inheritance in the context of
laterally uniform melt transport. In particular, it considers the
preservation or attenuation of trace-element heterogeneity during
simple, vertical melt transport and aggregation.

\subsection{Model overview}

In this paper we aim to understand the end-member scenario of
source-heterogeneity with laterally uniform melt transport. We ask the
questions: \textit{(i)} Which wavelengths of heterogeneity are
preserved in the magma during its segregation and ascent through an
upwelling, one-dimensional column of mantle rock?  \textit{(ii)} Which
wavelengths of heterogeneity are filtered out?  \textit{(iii)} How
does a trace element's partition coefficient affect the transfer of
the source heterogeneity through the melting region?  \textit{(iv)}
How does the low-productivity tail of deep, low-degree melting affect
the transfer from source? The model we develop to answer these
questions envisions one-dimensional, vertical, steady melt segregation
to the base of the crust.

Following \cite{depaolo96}, we regard mantle trace-element
heterogeneity as the sum of sinusoidal variations of different
wavelengths.  The peak and trough of the sinusoidal cycle reflects a
source that is incompatible-element ``enriched'' and ``depleted,''
respectively.  It is important to note, therefore, that the model does
not have a binary distinction between sources. Instead, the source
composition is continuous and smoothly varying.  We note that any
smooth and periodic function can be constructed by an appropriately
weighted sum of sinusoidal (Fourier) modes.

The chemistry described by the model is the partitioning and
segregation of fictive trace elements.  We define trace-element
equilibrium in the common way through partition coefficients, $\pco$,
such that in equilibrium, $\cs=\pco\cl$. The mass of trace elements is
conserved. Conservation statements alone, however, do not constrain
the way that elements are transferred between phases. A common
description of melting in the geochemical literature is as a
near-fractional process \citep[e.g.,][]{mckenzie91, yoder1976,
  johnson1990_jgr, beattie1993_nature, hellebrand2001_nature,
  asimow2004_jpet}, whereby incremental melts are in equilibrium with
the composition of their parental solid prior to their near complete
extraction. Here we adopt this approach and quantify its implications
for the inheritance of chemical variability from the heterogeneous
source.

Melting rate is an important model parameter, as it ultimately
controls segregation of liquid from solid within the melting region.
We model two idealised patterns of melting rate: ``dry'' mantle
melting, in which the rate is constant from the solidus intersection
to the base of the crust; and ``wet'' mantle melting, where a
low-productivity tail precedes an interval of nominally dry melting.
Consideration of these two regimes is motivated by the expectation
that the height of the partially molten region could be important. In
particular, attenuation of source heterogeneity should be promoted if
the melt region simultaneously contains multiple cycles of the source
heterogeneity.  Segregating melts of the chemically diverse
sources aggregate and mix, which pulls the composition of enriched or
depleted melts back toward the mean composition of the melting region.
The more cycles of heterogeneity in the melting region, the greater is
this regression to the mean.  Dry and wet melting represent shorter
and longer column lengths, respectively, and should thus behave
differently.

In this context, reaction between melts and solid could also play a
role in attenuating source heterogeneity.  To explore this
possibility, we assume that the aggregated melt can react with the
solid to move toward chemical equilibrium. The reaction rate at which
this occurs is proportional to the chemical deviation from
equilibrium, as defined by the partition coefficient $\pco$.  If
reaction is infinitely fast, the model describes batch melting and
equilibrium transport.  For finite values of the reaction rate,
considered in section~\ref{sec:reaction}, partial equilibration
occurs. For zero reaction rate, the model describes fractional melting
and disequilibrium transport; this combination is commonly assumed in
geochemical models and is the focus of this paper.

The final consideration for the model is how to describe the output
melt chemistry in relation to that of the source that is input at the
bottom of the melting region. We quantify the transfer of incoming
source heterogeneity to outgoing magma variability in terms of the
admittance $\admitl$, a concept developed further below.  In short, it
is analogous to the mean-normalised variance for trace elements from a
suite of basaltic lavas or melt inclusions.

\subsection{Previous melt models investigating the transport of source
  heterogeneity}

Many previously published studies employ column models that assume
porous magmatic ascent, with full or partial aggregation of the melts
produced at different depths.  This is the basis on which
\cite{mckenzie85} and \cite{navon87} developed theories for
trace-element transport, showing that equilibration between liquid and
solid phases leads to transport rates that depend on the partition
coefficient.  Near equilibrium, heterogeneities of incompatible trace
elements move at the chromatographic velocity, which is intermediate
between the liquid and solid velocities and depends on the partition
coefficient and melt fraction. Under idealised conditions (i.e.,
neglecting diffusion and dispersion), transport in equilibrium
preserves chemical heterogeneities at all wavelengths.
\cite{kenyon90} and \cite{depaolo96} found that dispersion causes
attenuation at very short wavelengths of heterogeneity.

\cite{navon87} recognised that a long diffusion time is potentially
required to equilibrate the melt with the interior of solid grains,
and that this will lead to a more substantial deviation from ideal,
chromatographic behaviour. Disequilibrium models that explicitly track
diffusion along the radii of representative, spherical grains were
developed to address this issue \citep{qin92, iwamori92,
  iwamori93a}. They show that the effective partition coefficient can
be significantly higher than the equilibrium value if transfer into
the melt is rate-limited by diffusion through the solid. Thus partial
equilibration essentially traps incompatible elements in the grain
interior \citep[see also][]{liang16}. However, all of these studies
focused on steady-state solutions, which precludes a treatment of
chemical heterogeneity of the source.

\cite{kenyon90} and subsequent papers \citep{kenyon93, kenyon98}
considered how disequilibrium transport could attenuate (or preserve)
fluctuations of trace-element concentration in ascending magma. Her
models idealise pores as narrow, vertical sheets of magma that are
interleaved with slabs of solid. Both magma sheets and solid slabs
have uniform width; melt ascent rate is constant. Chemical equilibrium
is imposed at the liquid--solid interface. The liquid is assumed
well-mixed in the across-pore direction with zero diffusion parallel
to the flow; transport through the solid is by horizontal diffusion
only. A sinusoidally varying concentration of trace elements,
representing melt derived from a heterogeneous source, is injected
into the bottom of the domain and modified by exchange with the
solid. There is no melting in the interior of the domain.

\cite{kenyon90} presents an analytical solution to this problem.  The
solution is discussed in terms of vertical transport rates and
attenuation of heterogeneity amplitude. Both are considered as a
function of oscillation frequency and pore width and spacing. The
solid diffusivity is held at $10^{-17}$~m$^2$/sec. Attenuation
increases with frequency, such that melt oscillations with periods of
order 1000~years are eliminated over less than a kilometre of
rise. For mantle upwelling at 3~cm/yr, this period corresponds to a
source wavelength of about 30~m. At the same upwelling rate, source
heterogeneity wavelengths of order 10~km give periods of $10^5$~yr.
In Kenyon's models, these longer-period oscillations attenuate over
tens to hundreds of kilometres of rise.

Key to the question of disequilibrium during melt transport is knowing
the trace-element diffusivities.  Rare-Earth element diffusivities
were measured by \cite{vanorman01} and found, in general, to be
significantly smaller than assumed by \cite{kenyon90}. This would
reduce the rate of melt equilibration with the solid and hence also
reduce the attenuation of heterogeneity amplitude. Kenyon's theory
would then predict preservation of shorter-period
oscillations. However, Kenyon's model [\citeyear{kenyon90, kenyon93,
  kenyon98}] neglects melting.  It is well known that melting
transfers trace-element mass to the liquid phase over a finite range
of melt fractions (which for decompression melting translates to a
depth interval).  This should logically play a role in the attenuation
of heterogeneity.

Melting is included by \cite{liu17} in a model of vertical,
disequilibrium transport of trace-element heterogeneities.  Their
analysis focuses on the stretching of isolated, non-interacting
trace-element anomalies.  The use of isolated heterogeneities makes it
difficult to generalise to a multiscale view of mantle
heterogeneity. \cite{liu17} concluded that smaller heterogeneities are
more easily attenuated during melt segregation. This is reinforced by
a more detailed paper by \cite{liang18} as well as by the results
presented below.

Here we focus on the transfer of heterogeneity from the mantle to the
magma by progressive melting. We show that attenuation dominantly
occurs by melt segregation during the initial (deepest) phase of
melting. Our model assumes equilibrium melting and and melt transport
without chemical equilibration between melt and solid.  Our key
finding is that melt transport attenuates chemical heterogeneity of
the upwelling mantle, depending on partitioning of the element
considered, its lengthscale of variation in the source mantle, and the
vertical structure of melting rate. This remains true for partial
chemical equilibration. In melts delivered to the crust, wavelengths
of order 1~km can be preserved only for the most incompatible
elements.

\subsection{Outline of manuscript}

The manuscript is arranged as follows. In section~\ref{sec:the-model}
we explain the domain, boundary conditions, and governing equations of
the column model.  In section~\ref{sec:results} we illustrate the
behaviour of the model for simple scenarios of dry melting (with
constant productivity) and wet melting (which adds a low productivity
tail). We develop a physical argument for attenuation of trace-element
heterogeneity. And we examine the consequences of reactive
equilibration of liquid and solid. Section~\ref{sec:observations}
compares three observational datasets from the literature with model
predictions in terms of the variance of concentration. Finally,
section~\ref{sec:discussion} discusses the model and its
limitations. We return to the question of whether observed variability
is a consequence of source heterogeneity or non-uniform melt
transport. We conclude that source heterogeneity cannot fully explain
the chemical diversity of basalts, and that variability of melt
transport (e.g., channelised flow) is required. 

\section{Model of trace-element transport}
\label{sec:the-model}

We consider a one-dimensional domain aligned with gravity --- a
melting column. The top of the column is located at $z=0$ and
represents the Moho; the bottom of the column is located at $z=z_0<0$,
where $\vert z_0\vert$ is the depth at which upwelling mantle begins
to melt and its porosity becomes non-zero. 

The boundary condition at the bottom of the column represents the
mantle composition as it upwells steadily into the domain at a rate
$W_0$. It has a mean, which is independent of time, and a sinusoidal
fluctuation that represents the introduction of source
heterogeneity. We can express this in terms of the complex expression
\begin{equation}
  \label{eq:boundary_condition}
  \cs_0(t) = \csm_0 + \csf_0\text{e}^{i\freq_0 t},
\end{equation}
where $\csm$ is the steady part of the mantle concentration and $\csf$
is the complex amplitude of the fluctuating part, and hence also
determines the phase-angle (recall Euler's formula,
$\text{e}^{i\freq_0 t} = \cos\freq_0 t + i\sin\freq_0 t$).  The
subscript $0$ indicates quantities at the bottom of the column.

The frequency of the fluctuating part of the boundary is
\begin{equation}
  \label{eq:bc_omega}
  \freq_0 = \frac{2\pi W_0}{\lambda_0},
\end{equation}
where $W_0>0$ is the mantle upwelling speed at the bottom of the
column and $\lambda_0$ is a wavelength of heterogeneity in the mantle
prior to the onset of melting. 

\subsection{Governing equations of trace-element transport}
\label{sec:governing-equations}

Conservation of mass equations governing trace-element evolution in
the solid (mantle, $s$) and liquid (magma, $\ell$) phases are
\begin{subequations}
  \label{eq:conservation_equations}
  \begin{align}
    \label{eq:conservation_equations_s}
    (1-\phi)\rho\Ldiff{s}{\cs}{t} &= -\left(\cg - \cs\right)\Gamma - \rxn,\\
    \label{eq:conservation_equations_l}
    \phi\rho\Ldiff{\ell}{\cl}{t}  &= +\left(\cg - \cl\right)\Gamma + \rxn,
  \end{align}
\end{subequations}
where $\text{D}_j/\text{D}t$ is a Lagrangian derivative following a
parcel of phase $j$ ($s$ or $\ell$), $\cg$ is the trace element
concentration in the instantaneously produced melt with infinitesimal
mass per unit volume $\Gamma\infd t$, and $\rxn$ is the rate of an
interphase mass-exchange reaction.  $\Gamma$ represents the melting
rate (kg/m$^3$/s); it is strictly positive in the models we consider,
but we defer any further specification until later in the manuscript.
Equations~\eqref{eq:conservation_equations} state that the rate of
change of trace-element concentration in a moving parcel of solid
mantle \eqref{eq:conservation_equations_s} or liquid magma
\eqref{eq:conservation_equations_l} is due to interphase transfer by
melting and by reactive exchange. Macroscopic diffusion and dispersion
of trace elements are neglected for both phases.

Fractional melting and linear kinetics are specified by
\begin{align}
  \label{eq:fractional_melting}
  \cg &= \cs/\pco,\\
  \label{eq:reaction_rate}
  \rxn &= \rr\left(\cs - \pco\cl\right),
\end{align}
where $\pco\equiv\left[\cs/\cl\right]^\text{eq}$ is a partition
coefficient representing the equilibrium ratio of solid to liquid
concentration, and $\rr$ is a kinetic coefficient with units
kg/m$^3$/s.  Equation~\eqref{eq:fractional_melting} states that the
instantaneously produced melt is in equilibrium with the entire solid
residue (there is no freezing in the model domain).
Equation~\eqref{eq:reaction_rate} states that the exchange of
trace-element mass between phases occurs at a rate that is linearly
proportional to the difference from equilibrium.  We take both $\pco$
and $\rr$ to be constant and uniform within any solution of
equations~\eqref{eq:conservation_equations} but explore their
parametric control using suites of solutions.



For $\rr\to\infty$, reaction eliminates even the smallest deviations
from trace-element equilibrium and hence the column produces batch
melts.  In contrast, for $\rr\to 0$, reaction makes no contribution;
fractional melts travel up the column but do not equilibrate with the
residue they traverse. In this case, the column produces aggregated
fractional melts.  Below we explore the model behaviour across this
range and determine how large or small $\rr$ must be to effectively
obtain these end-member regimes.

\subsection{Expansion into trace element means and fluctuations}
\label{sec:decomposition}

The full solution to the problem can be expanded into steady and
fluctuating parts \citep{liang08}. The steady part represents the mean
concentration as a function of depth for all time; the fluctuating
part represents the temporal oscillations associated with
heterogeneity.  The expansion is written
\begin{subequations}
  \label{eq:decomposition}
  \begin{align}
    \label{eq:decomposition_l}
    \cs(z,t) &= \csm(z) + \csf(z)\text{e}^{i\freq t}, \\
    \label{eq:decomposition_s}
    \cl(z,t) &= \clm(z) + \clf(z)\text{e}^{i\freq t},
  \end{align}
\end{subequations}
where the functions $\csf(z)$ and $\clf(z)$ are the complex amplitudes
of fluctuation, which depend only on depth.  It is important to note
that while mean concentrations must obey $\csm,\clm > 0$, the
fluctuations must oscillate about zero so as to have zero
time-mean. The amplitude of the fluctuations is small enough that the
full solid and liquid concentrations $\cs,\cl$ are always
positive. Only the real part of concentrations $\cs$ and $\cl$ are
physically relevant.

The time-dependence in \eqref{eq:decomposition} has been expressed in
terms of an oscillatory function with the same frequency for the
liquid and the solid. The assumption of this form stems from the
linearity of the equations; the frequency of the solution is locked to
the frequency of the forcing at the boundary,
eqn.~\eqref{eq:boundary_condition}.  Therefore $\freq = \freq_0$
uniformly and for both phases.

Moreover, because the governing
equations~\eqref{eq:conservation_equations} are linear, superposition
applies and we can solve for the mean and fluctuations
separately. Substituting~\eqref{eq:fractional_melting},
\eqref{eq:reaction_rate} and \eqref{eq:decomposition} into
\eqref{eq:conservation_equations} and requiring the mean terms to
balance gives
\begin{subequations}
  \label{eq:mean}
  \begin{align}
    \label{eq:mean_s}
    (1-\phi)\rho W\diff{\csm}{z}
    &= -\left(\csm/\pco - \csm\right)\Gamma - \rr\left(\csm-\pco\clm\right),\\
    \label{eq:mean_l}
    \phi\rho w\diff{\clm}{z} 
    &= +\left(\csm/\pco - \clm\right)\Gamma + \rr\left(\csm-\pco\clm\right).
  \end{align}
\end{subequations}
At the bottom of the column, the mean concentrations satisfy
$\csm(z=z_0) = \csm_0$ and $\clm(z=z_0) = \csm_0/\pco$. The system
\eqref{eq:mean} is a set of coupled, linear, ordinary differential
equations.

The equations for the fluctuating part of the solution are partial
differential equations, but they can be converted to complex ODEs by
applying the time derivatives in \eqref{eq:conservation_equations} to
the expansion in \eqref{eq:decomposition}.  This gives
\begin{subequations}
  \label{eq:fluctuations}
  \begin{align}
    \label{eq:fluctuations_s}
    (1-\phi)\rho\left(i\freq\csf + W\diff{\csf}{z}\right) 
    &= -\left(\csf/\pco - \csf\right)\Gamma - \rr\left(\csf-\pco\clf\right),\\
    \label{eq:fluctuations_l}
    \phi\rho\left(i\freq\clf + w\diff{\clf}{z}\right)  
    &= +\left(\csf/\pco - \clf\right)\Gamma + \rr\left(\csf-\pco\clf\right).
  \end{align}
\end{subequations}
At the bottom of the column, the fluctuation amplitudes satisfy the
fluctuating part of the boundary
condition~\eqref{eq:boundary_condition}.  In particular,
$\csf(z_0)=\csf_0$ and $\clf(z_0)=\csf_0/\pco$.

The variable that is most relevant for comparison with observations is
$\vert\clf(0)\vert$, the amplitude of fluctuation in the liquid at
$z=0$, the top of the melting column.  For any regime, this will be
linearly proportional to the amplitude of forcing,
$\vert\csf(z_0)\vert$.  Hence we define and study a pair of quantities
called \textit{admittance} (sometimes called the modulus of transfer),
\begin{equation}
  \label{eq:def_admittance} 
  \admits \equiv \frac{\left\vert\csf(z)\right\vert}
  {\left\vert\csf(z_0)\right\vert}, \qquad\qquad
  \admitl \equiv \frac{\left\vert\clf(z)\right\vert}
  {\left\vert\csf(z_0)\right\vert}.
\end{equation}
Admittance is a crucial concept in the analysis presented here. It
represents the fraction of the column-bottom heterogeneity that is
present at some height in the column.  In other words, it is the part
of the signal that has not been attenuated at that height.

We will be particularly interested in the liquid admittance as a
function of the heterogeneity wavelength $\lambda_0$, given the
parameters $\pco$ and $\rr$. This is written as
$\admitl(\lambda_0\vert\pco,\rr)$, where the vertical line separates
the independent variable, wavelength, from the problem parameters,
partition coefficient and reaction-rate constant. Although the
admittances are defined at any height $z-z_0$ in the column, in this
manuscript they will be evaluated and plotted at the top of the column
($z=0$) unless otherwise specified.

\section{Analysis of melting columns}
\label{sec:results}

Upwelling and melt production in the melting column is written in
terms of equations for conservation of mass and momentum for two
interpenetrating fluids, a liquid phase (the magma) and a creeping
solid phase (the mantle) \citep{mckenzie84}. Assuming that compaction
stresses are negligible \citep{ribe85a, spiegelman93a}, the
one-dimensional expression of these equations can be written
\begin{subequations}
  \label{eq:meltcol-solutions}
  \begin{align}
    \label{eq:meltcol-solutions-phi}
    \phi + \phi_0\frac{w_0}{W_0}
    \left(\frac{\phi}{\phi_0}\right)^n
         &\approx  F\qquad\text{for }\phi\ll1, \\
    \label{eq:meltcol-solutions-w}
    w &= W_0\frac{F}{\phi},\\
    \label{eq:meltcol-solutions-W}
    W &= W_0\frac{1-F}{1-\phi},
  \end{align}
\end{subequations}
where $\phi,\,F,\,w$ and $W$ are all functions of $z$.  This solution
arises when permeability is related to porosity as
$k_\phi = k_0(\phi/\phi_0)^n$, where $k_0$ is the permeability at
reference porosity $\phi_0$ and $n$ is a constant
\citep[e.g.,][]{vonbargen86, miller14,
  rudge18}. In~\eqref{eq:meltcol-solutions-phi},
\begin{equation}
  \label{eq:characteristic_melt_speed}
  w_0 = \frac{k_0\Delta\rho g}{\phi_0\mu}
\end{equation}
is a characteristic, buoyancy-driven melt speed for magma buoyancy
$\Delta\rho g$ and viscosity $\mu$. Uncertainty in the appropriate
value of $k_0$ for the mantle translates to uncertainty in the rate of
melt segregation. Unless otherwise specified, we use
$k_0=10^{-12}$~m$^2$ and $n=2$ in this paper. The degree of melting is
denoted by $F(z)$ and can be computed from a known melting rate
$\Gamma(z)$ as $F(z) = \int_{z_0}^{z}\Gamma(z)/\rho W_0\,\infd
z$. Further details are provided in Appendix~\ref{sec:melting-cols}.

We consider two simplified melting scenarios and their consequences
for filtration of mantle heterogeneity.  The first is a ``dry''
scenario, where melting begins at about 70~km depth and proceeds with
constant isentropic productivity to the surface.  The second is a
``wet'' scenario, where melting begins at about 120~km depth with the
production of volatile-rich melts at very low productivity;
productivity then increases with ascent above 70~km. Both columns
reach a total degree of melting of 23\%.

In sections~\ref{sec:admittance-dry} and \ref{sec:admittance-wet},
below, we present results from the dry and wet scenarios. These are
obtained by solving eqns.~\eqref{eq:mean} and \eqref{eq:fluctuations}
with no reaction ($\rr=0$), representing disequilibrium transport of
aggregated fractional melts.  The most important characteristics of
the results are described and illustrated.  All of these
characteristics can be explained within a simple, unified theory,
which is provided in section~\ref{sec:admittance-theory}.  With this
theory for disequilibrium transport in place, we then revisit the dry
and wet melting columns with partial equilibration ($\rr>0$) in
section \ref{sec:reaction}.

\subsection{Dry column: constant melt productivity}
\label{sec:admittance-dry}

The model assumes a melting rate driven by decompression, with a
uniform isentropic productivity $\iprod\equiv\Fmax/z_0$. The melting
rate is then
\begin{equation}
  \label{eq:gamma_uniform}
  \Gamma = \rho W_0 \iprod.
\end{equation}
and hence the degree of melting, $F(z) = \iprod(z-z_0)$, is linear
with height in the column.  The resulting column model is illustrated
in Appendix~\ref{sec:simplest} for a case with
$\Fmax=\iprod \vert z_0\vert = 0.23$. See
appendices~\ref{sec:melting-cols} and \ref{sec:simplest} for further
details.

The solution $\phi(z)$, obtained analytically from
equation~\eqref{eq:meltcol-solutions-phi} when $n=2$, can be
substituted into \eqref{eq:meltcol-solutions-W} and both of these into
equation~\eqref{eq:fluctuations_s} for the fluctuations in the solid
phase.  Under disequilibrium melt transport ($\rr=0$), this equation
can be solved analytically (Appendix~\ref{sec:simplest}) to give the
solid admittance as 
\begin{subequations}
  \begin{align}
    \label{eq:solid-admittance-exact}
    \admits &= \left(1-F\right)^{1/\pco-1} \\
    \label{eq:solid-admittance-approx}
            &\approx \e^{-F/\pco} = \e^{-(z-z_0)/\lambda_T}.
  \end{align}
\end{subequations}
The exact result \eqref{eq:solid-admittance-exact} is identical to the
well-known fractional melting solution of~\eqref{eq:mean_s} for the
mean concentration in the residue \citep{shaw06}.  The approximation
\eqref{eq:solid-admittance-approx} is valid for incompatible elements
at small degrees of melting.  It shows that the attenuation of
fluctuations occurs over a melting interval $F \lesssim \pco$.  We
refer to this interval as the \textit{transfer regime} because it
represents the region in which most of the trace element is
transferred from the solid to the liquid. The height of the transfer
regime $\lambda_T$ becomes the characteristic lengthscale for the
attenuation of chemical variability. For constant isentropic
productivity $\iprod$,
\begin{equation}
  \label{eq:define-transfer-height}
  \lambda_T = \pco/\iprod.
\end{equation}
The transfer regime will be important in understanding the admittance
of trace elements in the liquid phase.

Equation~\eqref{eq:fluctuations_l} governing trace-element
fluctuations in the liquid phase does not have a fully general,
analytical solution.  However, we derive an analytical bound on the
admittance
\begin{equation}
  \label{eq:bound_liquid_admittance}
  \admitl \le \clm/\csm_0
\end{equation}
in Appendix~\ref{sec:simplest}. This inequality states that the
admittance of the liquid phase can be no larger than the ratio of the
mean liquid concentration to the mean source composition.  In other
words, for the liquid phase, heterogeneity is attenuated at least as
fast as the mean concentration is diluted.

\begin{figure}[ht]
  \centering
  \includegraphics[width=\textwidth]{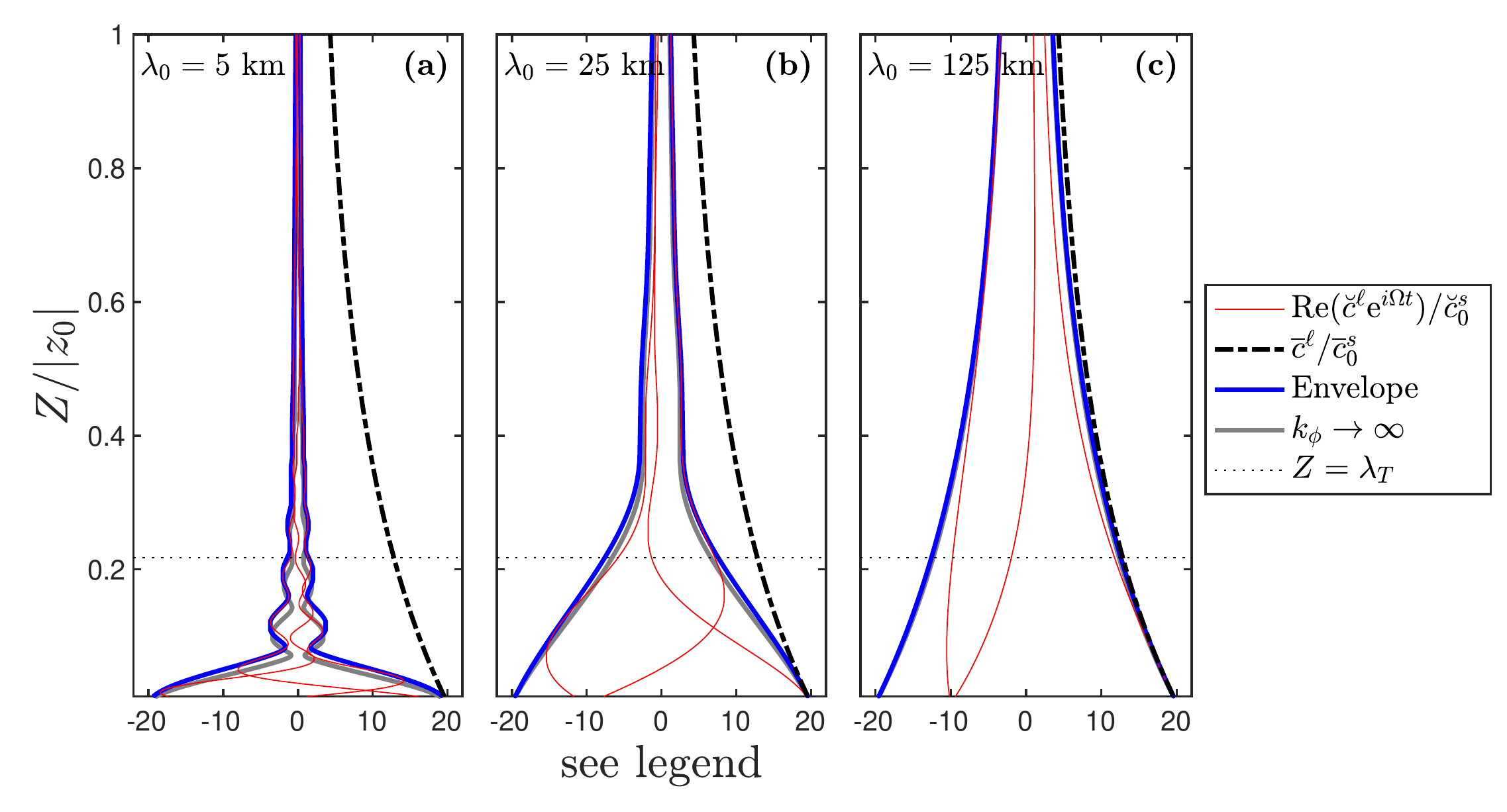}
  \caption{The vertical concentration structure of a trace element
    with $\pco=0.05$ in a column of height $\vert z_0\vert=70$~km,
    with uniform melt productivity and $\Fmax=0.23$. The
    transfer-regime height is
    $\lambda_T=D\vert z_0\vert/\Fmax \approx 15$~km. Curves show the
    mean $\clm$ (black) and fluctuations
    $\real\left(\clf \text{e}^{i\freq t}\right)$ (red), normalised by
    the associated value in the unmelted mantle source (see legend for
    details). The wavelength $\lambda$ of the input heterogeneity is
    \textbf{(a)} 5~km, \textbf{(b)} 25~km and \textbf{(c)} 125~km. In
    each panel, three lines are plotted for
    $\real\left(\clf \text{e}^{i\freq t}\right)$ evaluated at three
    different times by solving equations~\eqref{eq:fluctuations}
    numerically. Blue lines show the envelope for all possible
    times. Grey lines show the infinite-permeability asymptotic model
    of equation~\eqref{eq:infinite-k-admittance}. Details of the
    physical model for the melting column are given in
    Appendix~\ref{sec:simplest}.}
  \label{fig:trace_columnview}
\end{figure}

Numerical solutions to equations~\eqref{eq:fluctuations} are obtained
using Runge-Kutta methods.  Figure~\ref{fig:trace_columnview} shows
numerical solutions of trace-element concentrations in the liquid as a
function of height $Z = z-z_0$ in the column. The fluctuations are
plotted at three different times (red lines) by computing the real
part of~\eqref{eq:decomposition_l},
$\real\left[\clf(z)\text{e}^{i\freq t}\right]$. The envelope of the
liquid fluctuations (blue lines) is given by the modulus of the
fluctuation amplitude $\left\vert\clf(z)\right\vert$.  All of these
curves represent an incompatible element with $\pco=0.05$.

The three panels of Figure~\ref{fig:trace_columnview} show results for
three wavelengths of heterogeneity, $\lambda_0=5$, $25$ and
$125$~km. Shorter wavelengths are more efficiently attenuated by the
column than longer wavelengths. Indeed, the fluctuations of the
$\lambda_0=5$~km case (panel~(a)) are qualitatively eliminated. Note
that as predicted in equation~\eqref{eq:bound_liquid_admittance}, the
envelope of fluctuations remains within the bound defined by the mean
concentration. As the wavelength $\lambda_0\to\infty$, the envelope
converges to the mean concentration.

We can understand the envelope structure in
Figure~\ref{fig:trace_columnview} through an asymptotic analysis of
the governing equation~\eqref{eq:fluctuations_l} (see
appendix~\ref{sec:simplest} for details).  When the permeability is
taken to be infinite, upwelling of the liquid is much faster than that
of the solid.  In this limit (and for $\pco, F \ll 1$), an asymptotic
admittance can be computed exactly
\begin{equation}
  \label{eq:infinite-k-admittance}
  \admitl\sim \frac{\sqrt{1 + \e^{-2Z/\lambda_T} -
      2\e^{-Z/\lambda_T}\cos(2\pi Z/\lambda_0)}}
    {F\sqrt{1 + (2 \pi \lambda_T/\lambda_0)^2}}.
\end{equation}
This function is plotted in Figure~\ref{fig:trace_columnview} as grey
lines that closely match the envelope obtained numerically.  The gross
decay of amplitude is controlled by the denominator of
\eqref{eq:infinite-k-admittance}; the envelope fluctuations are
controlled by the numerator.  We consider each of these in turn.

For sufficiently small partition coefficient $\pco$ we have
$Z\gg\lambda_T$ near the top of the column.  In this case, the
numerator of \eqref{eq:infinite-k-admittance} is
$\sim 1$ and we have
\begin{equation}
  \label{eq:infinite-k-column-top}
  \admitl\sim \frac{1}{F\sqrt{1 + (2 \pi \lambda_T/\lambda_0)^2}}\qquad
  \text{for }Z\gg\lambda_T.
\end{equation}
Recall that $\lambda_0$ is the wavelength of mantle heterogeneity in
the source mantle.  This equation indicates that near the top of the
column, there are two admittance regimes.  The first regime has
$\lambda_0\gg\lambda_T$ and hence $\admitl \sim F^{-1}$, independent
of $\pco$ and $\lambda_0$.  This behaviour is achieved for highly
incompatible elements and/or for large heterogeneity wavelength. All
source heterogeneity is mirrored in the melt and hence this is an
upper bound on the admittance over parameter space.  The second regime
has $\lambda_T\gg\lambda_0$ and hence
$\admitl \sim \lambda_0\iprod/(2\pi F \pco)$. Admittance thus
decreases with partition coefficient and increases with wavelength and
melt productivity.

\begin{figure}[ht]
  \centering
  \includegraphics[width=\textwidth]{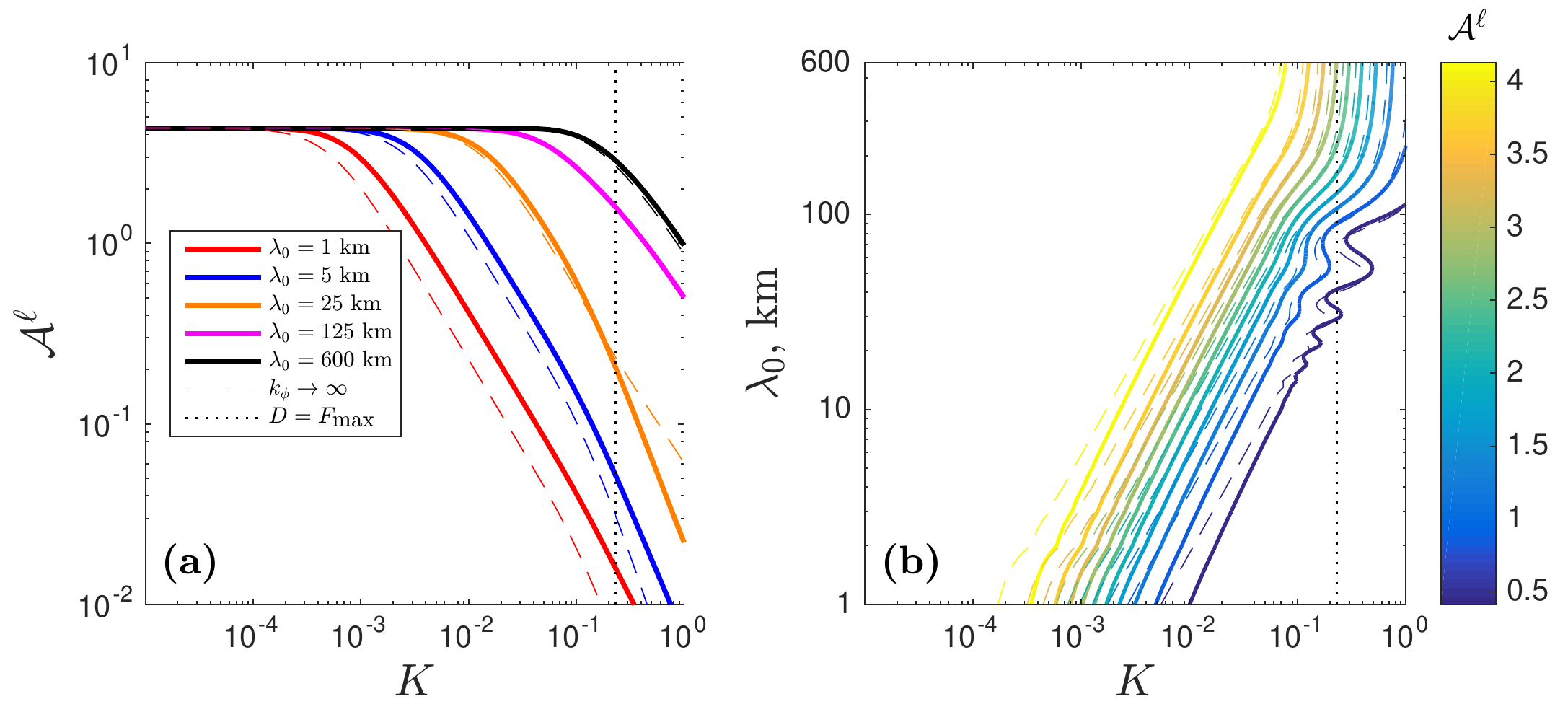}
  \caption{Admittance $\admitl$ of trace-element heterogeneity in the
    dry melting column (Fig.~\ref{fig:trace_columnview}) with maximum
    degree of melting $\Fmax$ of 23\%. Solid lines are obtained from
    numerical integration of equation~\eqref{eq:fluctuations_l};
    dashed lines are calculated with the asymptotic
    solution~\eqref{eq:infinite-k-admittance}. \textbf{(a)} $\admitl$
    as a function of partition coefficient $\pco$ for various
    wavelengths of heterogeneity, as in legend. \textbf{(b)} Contours
    of constant $\admitl$ as a function of $\pco$ and input
    heterogeneity wavelength $\lambda_0$.  Other parameters as in
    Fig.~\ref{fig:trace_columnview}}
  \label{fig:admittance_dry}
\end{figure}

Further down in the column, where $Z/\lambda_T$ is $O(1)$, the
numerator of \eqref{eq:infinite-k-admittance} plays a
role. Oscillations in the envelope occur at the source-heterogeneity
wavelength $\lambda_0$, but their amplitude decays over the
transfer-regime lengthscale.  In the limit of $Z\to 0$, we can
approximate the exponential and cosine functions with Taylor series
and simplify to leading order to give $\admitl\sim 1/\pco$.  Hence we
note that the asymptotic behaviour of admittance is closely related to
the canonical fractional melting model at the top
($\clm/\csm_0\sim F^{-1}$) and bottom ($\clm/\csm_0\sim \pco^{-1}$) of
the column.

Figure~\ref{fig:admittance_dry} summarises column-model results for a
range of heterogeneity wavelength and partition coefficient, in terms
of the liquid admittance at the top of the column $\admitl(z=0)$. The
two panels are different ways of visualising the same information: the
filtration properties of the melting column.  Panel~(a) displays the
two regimes that are identified by the infinite permeability model in
equation~\eqref{eq:infinite-k-column-top}. At small $\pco$, we are in
the regime where $\lambda_0\gg\lambda_T$ and hence where
$\admitl(0) \sim \Fmax^{-1}$. The column-top admittance in this regime
is independent of wavelength. At large $\pco$, we are in the other
asymptotic regime that has $\admitl(0) \propto \lambda_0/\pco$.
Considering the full range of $\pco$ in panel~(a), we note that
heterogeneity at a 1~km wavelength is severely attenuated by transport
through the column, except at the lowest partition coefficients (e.g.,
Barium, $\pco\approx 10^{-4}$).  In contrast, heterogeneity at a 125~km
wavelength is generally preserved in the column-top aggregated melts.

Panel~(b) of Figure~\ref{fig:admittance_dry} shows the same numerical
results, plotted in terms of contours of equal $\admitl$ in a
wavelength--partition coefficient space. The thin, dashed lines are
contours of the infinite-permeability
model~\eqref{eq:infinite-k-admittance}, evaluated at the column
top. In the upper-left region of this plot, where both the column
height and the heterogeneity wavelength are much greater than the
transfer regime ($Z_\text{max},\lambda_0\gg\lambda_T$), admittance is
uniformly high ($\admitl\sim\Fmax^{-1}$) and heterogeneity is
preserved. Moving from this region to the right takes us toward the
regime where $\lambda_T\gg\lambda_0$.  To leading order, admittance in
this regime varies as $\admitl \propto \lambda_0/\pco$ (hence the
contours have a slope $\sim 1$).

In Figure~\ref{fig:admittance_dry}(b), the oscillations in admittance
near $\pco=\Fmax$ arise from the sinusoidal term in
equation~\eqref{eq:infinite-k-admittance}.  The deviations from the
overall trend are small, however, and occur only when the admittance
is already low.  Hence the systematics of $\admitl$ as a function of
heterogeneity wavelength and partition coefficient is well-described
by equation~\eqref{eq:infinite-k-column-top}. This equation rests on
the assumptions of rapid melt segregation and a column that is much
taller than the transfer regime.  A more physically detailed
explanation for the systematics of admittance is provided in
section~\ref{sec:admittance-theory}, below.

\subsection{Wet column: variable melt productivity due to volatiles}
\label{sec:admittance-wet}

We next consider a melting column model with a mantle source that
contains volatiles (e.g., water and carbon). Although these volatiles
are present in small concentration, they drastically lower the solidus
temperature at a given pressure
\citep[e.g.,][]{dasgupta06a}. Therefore, melting begins at a higher
pressure. More importantly, the degree of melting $F$, does not
increase linearly with height in the column, as it did in the dry
column model. The melting rate can still be described as
in~\eqref{eq:gamma_uniform}, but the productivity $\iprod$ is no
longer constant; it now depends on $z$ and so we replace it with
$\infd{F}/\infd{z}$, which is a function of $z$.  The
zero-compaction-length column solution is given
by~\eqref{eq:meltcol-solutions}, but with a nonlinear $F(z)$. Details
of this model are given in Appendix~\ref{sec:meltcol-volatiles} (and
Fig.~\ref{fig:dry_wet_melting_col}). In the present treatment, the
volatile is taken to be water with a partition coefficient
$\pco_w = 0.01$. Melting proceeds to the same final extent, however:
$\Fmax = 0.23$. In Appendix~\ref{sec:meltcol-volatiles}, a simple
thermochemical model is introduced, where $F$ is expressed as a
function of temperature and $T(z)$ is obtained by numerical solution
of an energy conservation equation.

\begin{figure}[ht]
  \centering
  \includegraphics[width=\textwidth]{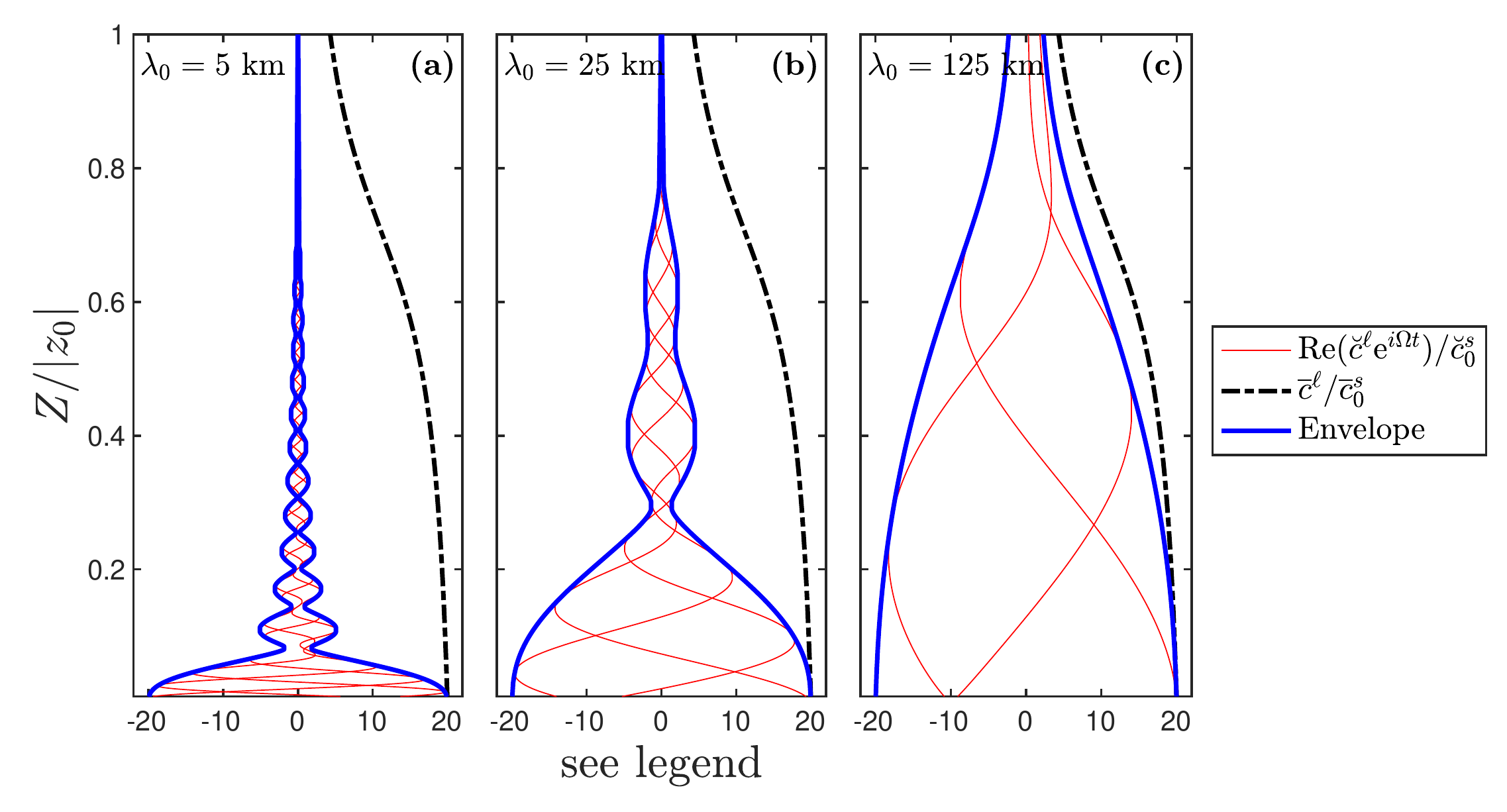}
  \caption{Melting column models with wet mantle source containing
    100~ppm water. Panels and lines as in
    Figure~\ref{fig:trace_columnview}. The onset of wet melting is at
    120~km depth and proceeds with non-uniform isentropic
    productivity. Details of the melting column physical and
    thermo-chemical models are given in
    Appendix~\ref{sec:meltcol-volatiles} and
    Figure~\ref{fig:dry_wet_melting_col}.}
  \label{fig:trace_columnview_vol}
\end{figure}

Figure~\ref{fig:trace_columnview_vol}, as for
Fig.~\ref{fig:trace_columnview}, displays solutions for $\clm(z)$ and
$\clf(z)$ for a trace element with $\pco=0.05$. The trace-element
concentrations in the liquid phase are plotted as a function of height
$Z = z-z_0$ in the column (with $z_0=-120$~km).  The mean (black line)
is separated from the fluctuations (red lines), which have an envelope
given by the blue lines.  The fluctuating part is computed at three
different times. From these curves we can draw a similar conclusion as
in section~\ref{sec:admittance-dry}.  Shorter-wavelengths fluctuations
are more efficiently filtered by the melting column than longer
wavelengths. The envelope of fluctuations remains within the mean
concentration, in agreement with the analytically derived bound in
eqn.~\eqref{eq:bound_liquid_admittance}, which was obtained for the
dry model.  Moreover, as the wavelength $\lambda_0\to\infty$, the
envelope converges to the mean concentration.

The wet column model has an onset of melting that is much deeper:
120~km versus 70~km for the dry case.  It also has a non-constant
productivity of isentropic decompression $\infd F/\infd z$; indeed,
there is a low-productivity ``tail'' at depths below about 60~km.  The
depth axis is normalised by the column height in
Fig.~\ref{fig:trace_columnview_vol}, so a direct comparison to depths
in Fig.~\ref{fig:trace_columnview} is not straightforward.  But it is
clear that the black curves showing the canonical fractional melting
solution differ between the wet and dry columns.  A larger
height-fraction of the wet column has low $F$ and hence high
$\clm/\csm_0$.  The envelope for the fluctuating part of the trace
element concentration (blue curve), however, diverges from its upper
bound deeper in the wet column than in the dry column --- both in the
relative terms of the fractional height as well as in the absolute
depth.

\begin{figure}[ht]
  \centering
  \includegraphics[width=\textwidth]{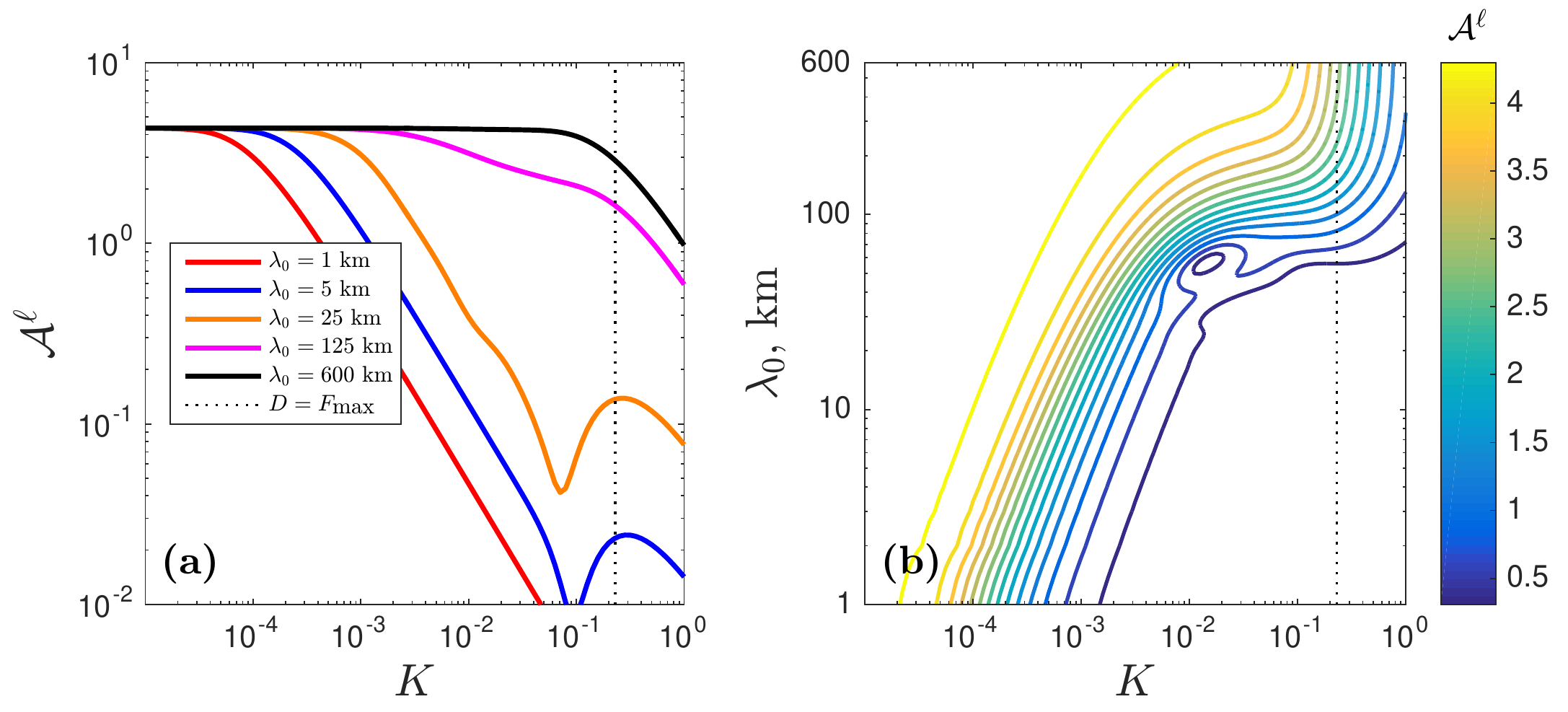}
  \caption{Admittance $\admitl$ of trace-element heterogeneity in the
    wet melting column with maximum degree of melting $\Fmax$ of
    23\%. Panels and parameters as in Fig.~\ref{fig:admittance_dry}
    except for the source water content of 100~ppm.}
  \label{fig:admittance_wet}
\end{figure}

Figure~\ref{fig:admittance_wet}, as in Fig.~\ref{fig:admittance_dry},
summarises the behaviour of the admittance for a suite of wet column
model calculations. $\admitl$ is plotted as a function of mantle
heterogeneity wavelength $\lambda_0$ and partition coefficient
$\pco$. The general trend for the wet columns is the same as for the
dry model: heterogeneity is transported to the surface with less loss
of amplitude when $\pco$ is small and when $\lambda_0$ is large.

However, comparing Figures~\ref{fig:admittance_wet} and
\ref{fig:admittance_dry} in more detail, there are significant
differences in the admittance structure. Lines in
Fig.~\ref{fig:admittance_wet}(a) show a more pronounced drop-off when
compared with Fig.~\ref{fig:admittance_dry} (except the black line)
and correspondingly, in \ref{fig:admittance_wet}(b), the contours
shift leftward. Both panels indicate that the liquid admittance
$\admitl$ becomes smaller with the existence of volatiles. In other
words, volatiles enhance the attenuation of mantle heterogeneity.

\begin{figure}[ht]
  \centering
  \includegraphics[width=0.5\textwidth]{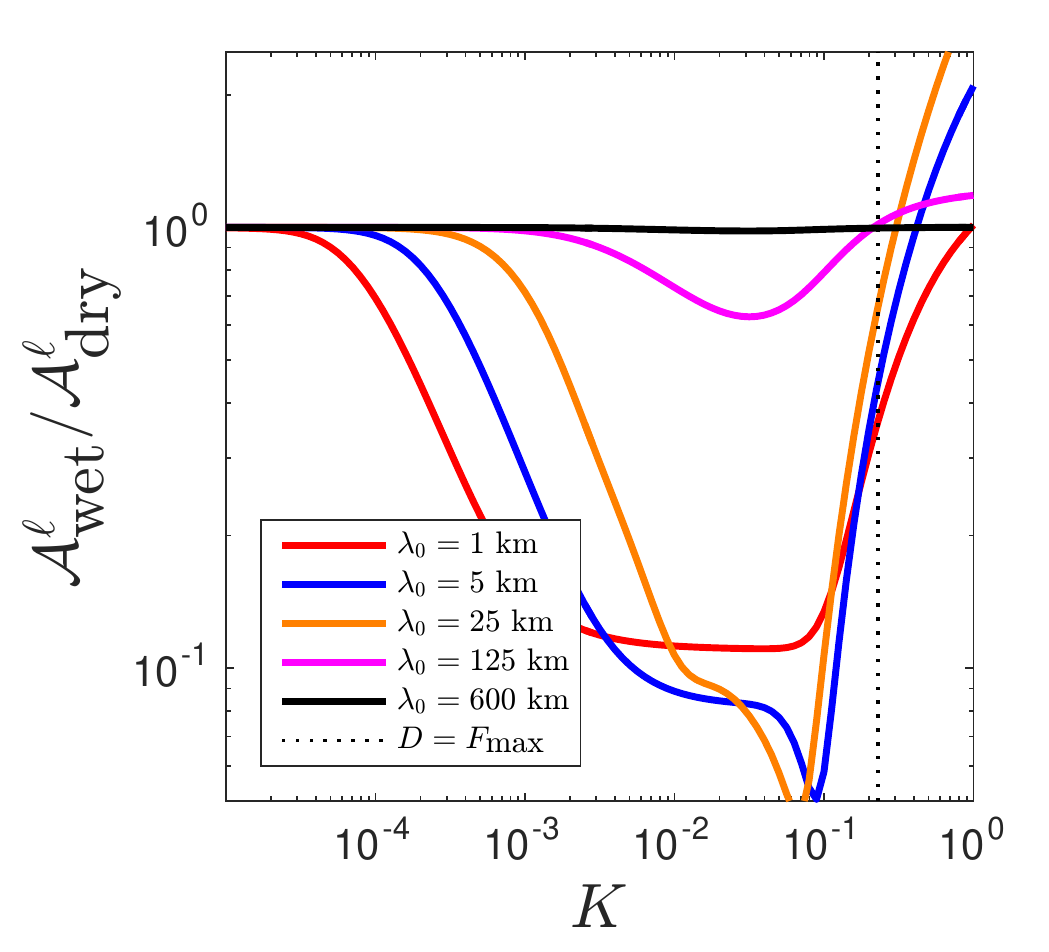}
  \caption{Ratio of admittance in the volatile model
    $\admitl_\text{wet}$ to admittance in the simple model
    $\admitl_\text{dry}$ as a function of partition coefficient $\pco$
    for various wavelengths of heterogeneity (see legend).  Parameters
    are the same as those in Fig.~\ref{fig:admittance_dry} and
    Fig.~\ref{fig:admittance_wet}}
  \label{fig:admittance_dry_wet}
\end{figure}

This enhanced attenuation is certainly evident when $\pco\ll \Fmax$.
However, for partition coefficients that approach $\Fmax$,
non-monotonic behaviour appears in the curves of $\admitl$
(Fig.~\ref{fig:admittance_wet}(a)). In
Figure~\ref{fig:admittance_dry_wet}, a plot of the ratio of admittance
in the wet and dry cases $\admitl_\text{wet}/\admitl_\text{dry}$
highlights this behaviour. Where the wet/dry admittance ratio is less
than unity, the wet column is more attenuating than the dry
column. The ratio increases toward unity as $\pco\to \Fmax$ from below
and, for some wavelengths, even exceeds unity.  The black line, for a
wavelength of heterogeneity of $\lambda_0=600$~km, shows that at
sufficiently long wavelength, the filtration effects of wet and dry
columns are indistinguishable.

There are other irregularities of the curves in
Figures~\ref{fig:admittance_wet} and \ref{fig:admittance_dry_wet}.
These generally occur when $\admitl$ is already significantly less
than unity, so they are of no practical importance and are not
discussed further. 

Above we have described results for trace element transport and
filtration of heterogeneity signals in dry and wet melting columns.
The most salient features have been highlighted but no explanation was
provided.  In the next section, we explain all of these results within
a single conceptual and quantitative framework.  This framework may be
usefully applied beyond the simple, one-dimensional models presented
here. 

\subsection{A simplified theory of wavelength selection}
\label{sec:admittance-theory}

\begin{figure}[h!]
  \centering
  \includegraphics[width=\textwidth]{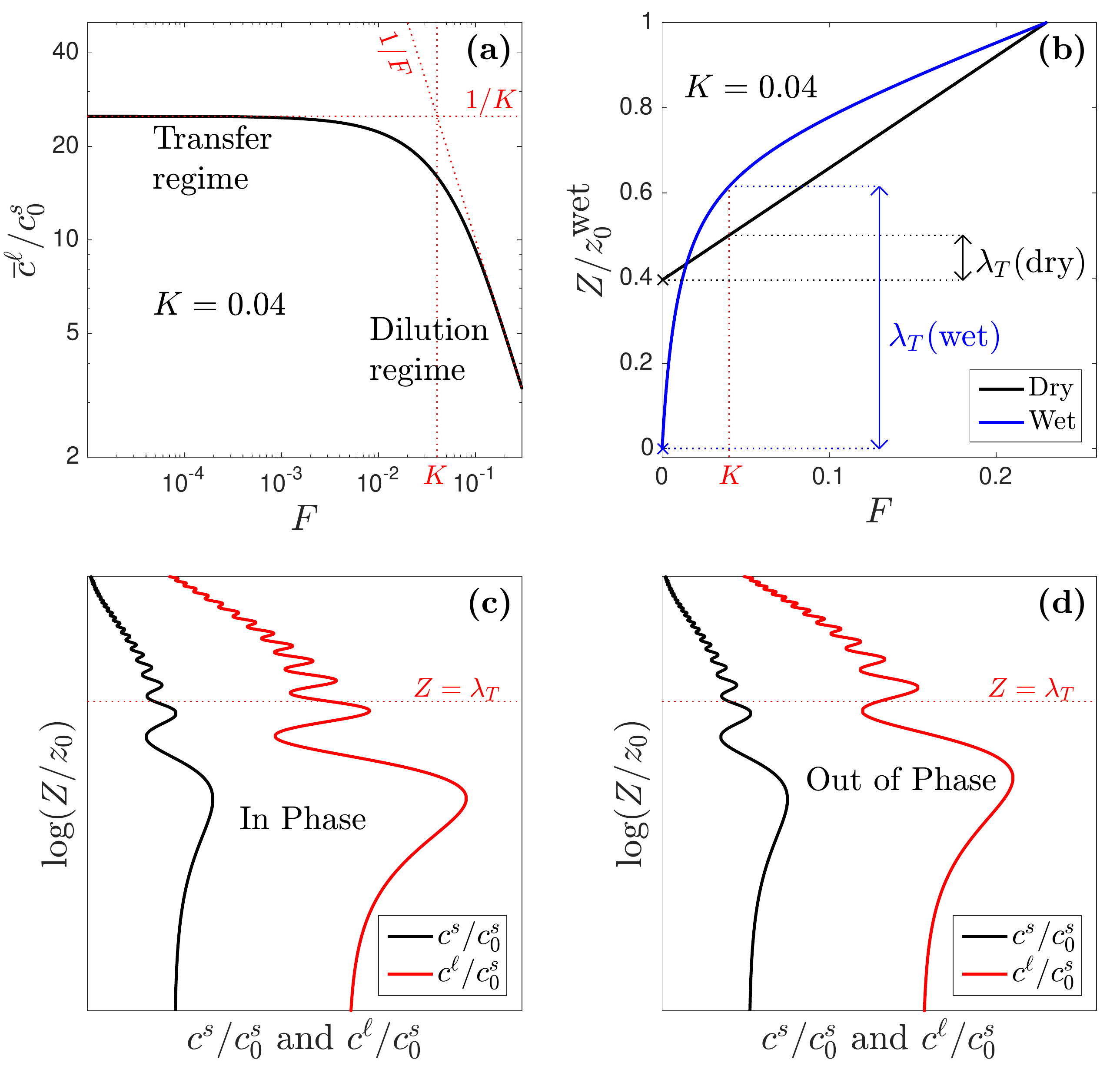}
  \caption{Plots to illustrate the mechanism of attenuation.
    \textbf{(a)} The canonical model of fractional melting
    ($\pco=0.04$), plotted in a log-log space.  The red line at
    $F=\pco$ delimits the transfer regime and the dilution regime. The
    liquid concentration approaches $1/\pco$ in the transfer regime,
    whereas it approaches $1/F$ in the dilution regime. \textbf{(b)}
    The length of transfer regime in the simple model and the volatile
    model. Black lines show the dry case; blue lines show the wet
    case. The red line denotes $F=\pco$ with $\pco=0.04$. Bottom
    panels are schematic diagrams showing how solid and liquid
    concentration can be ``in phase'' \textbf{(c)} or ``out of phase''
    \textbf{(d)} in the transfer regime.  Red lines represent the
    liquid phase; black lines represent the solid phase. Horizontal
    dotted lines mark $Z = \lambda_T$, the upper boundary of the
    transfer regime, where $F=\pco$.}
  \label{fig:Transfer_Regime}
\end{figure}

For any trace element with a fixed value of $\pco$, the vertical
evolution of an aggregated fractional melt has two regimes: one at
depths where $F(z)<\pco$, and one where
$F(z)>\pco$. Figure~\ref{fig:Transfer_Regime}(a) shows that there is a
significant change in trace-element variation with $F$ across this
boundary. In the transfer regime, incremental melts transfer
trace-element mass from the solid to the liquid, keeping the liquid
concentration nearly constant.  In the dilution regime, the solid is
depleted and incremental melts only dilute the concentration of the
liquid.  These two regimes map onto the steady, one-dimensional
melting column because at any depth (and corresponding $F$), the mean
liquid concentration is equal to the closed-system, aggregated melt of
the mean initial source concentration.

In a melting column, the transfer regime occurs toward the bottom,
where $F(z)<\pco$, and the dilution regime holds toward the top, where
$F(z)>\pco$. Trace-element source heterogeneity is transferred into
the liquid in the transfer regime and gets diluted in the dilution
regime. For elements with $\pco \ll \Fmax$, dilution affects the
admittance $\admitl$ uniformly; this creates the upper bound on
$\admitl$ in Figures~\ref{fig:admittance_dry} and
\ref{fig:admittance_wet}. Elements with $\pco \gtrsim \Fmax$ are
incompletely transferred to the liquid phase and hence their $\admitl$
never reaches the upper bound of $1/\Fmax$.

However, at a fixed $\pco \ll \Fmax$,
Figures~\ref{fig:admittance_dry} and \ref{fig:admittance_wet} show
that smaller wavelength of heterogeneity $\lambda_0$ is associated
with smaller $\admitl$. This additional attenuation cannot take place
in the dilution regime because melting of the depleted solid there
dilutes trace elements independent of their wavelength of variation.

Attenuation of trace-element variations in the liquid can occur in the
transfer regime, where the solid retains a significant fraction of the
total amount of trace element.  Then the difference in the phase-angle
of oscillation between the liquid and solid causes the attenuation.
If spatial variations in the liquid and solid remain \textit{in
  phase}, then additional fractional melting increases the variability
of the liquid; this is shown in Figure~\ref{fig:Transfer_Regime}(c).
If the spatial variations go \textit{out of phase}, as shown in
panel~(d), then fractional melting transfers higher-than-average
concentrations where the aggregated melt has a lower-than-average
concentration (and vice versa). This reduces variability in the liquid
phase.  Hence it is phase differences within the transfer regime that
cause attenuation of trace-element variability and reduce $\admitl$.

At the bottom of the melting column, where $F=0$, the solid and liquid
concentrations are in phase.  Previously we defined the height
$\lambda_T$ of the transfer regime as the interval of $z$ over which
$F$ ranges from 0 to $\pco$. Figure~\ref{fig:Transfer_Regime}(b) shows
how $\lambda_T$ is defined for dry and wet models for a given $\pco$.
A phase shift arises within this height interval if the melt and solid
travel at different speeds. Furthermore, if the wavelength of
heterogeneity is small compared to $\lambda_T$ then it is easier for a
speed difference (i.e., for melt segregation) to cause a phase shift.
The amount of attenuation, and hence the reduction in admittance,
should scale with the average difference of phase-angle between the
liquid and the solid.

This can be clarified by considering the real part of the integrand in
the expression for the liquid
admittance~\eqref{eq:simple_admittance_l_integral}. Although the full
equation is more complicated, its essence is evident in this term. It
is also helpful to make the approximation
$\left(1-F\right)^{1/\pco - 1} \approx \e^{-F/\pco}$ to give
\begin{equation}
  \label{eq:attenuation_schematic}
  \e^{-F/\pco}\cos\left[\freq\left(t^s - t^\ell\right)\right].  
\end{equation}
This expression has two parts.  The exponential part represents the
mean transfer of concentration from the solid to the liquid; it
highlights the characteristic melting scale over which the solid
becomes depleted.  The cosine term represents the effect of
phase-angle difference between the solid and liquid.  In particular,
$t^s - t^\ell \equiv \Delta t(F)$ is the difference in transit time
for the solid and the liquid to travel from the bottom of the melting
column to the height $Z$, at which the degree of melting is $F$.

\begin{figure}[ht]
  \centering
  \includegraphics[width=0.8\textwidth]{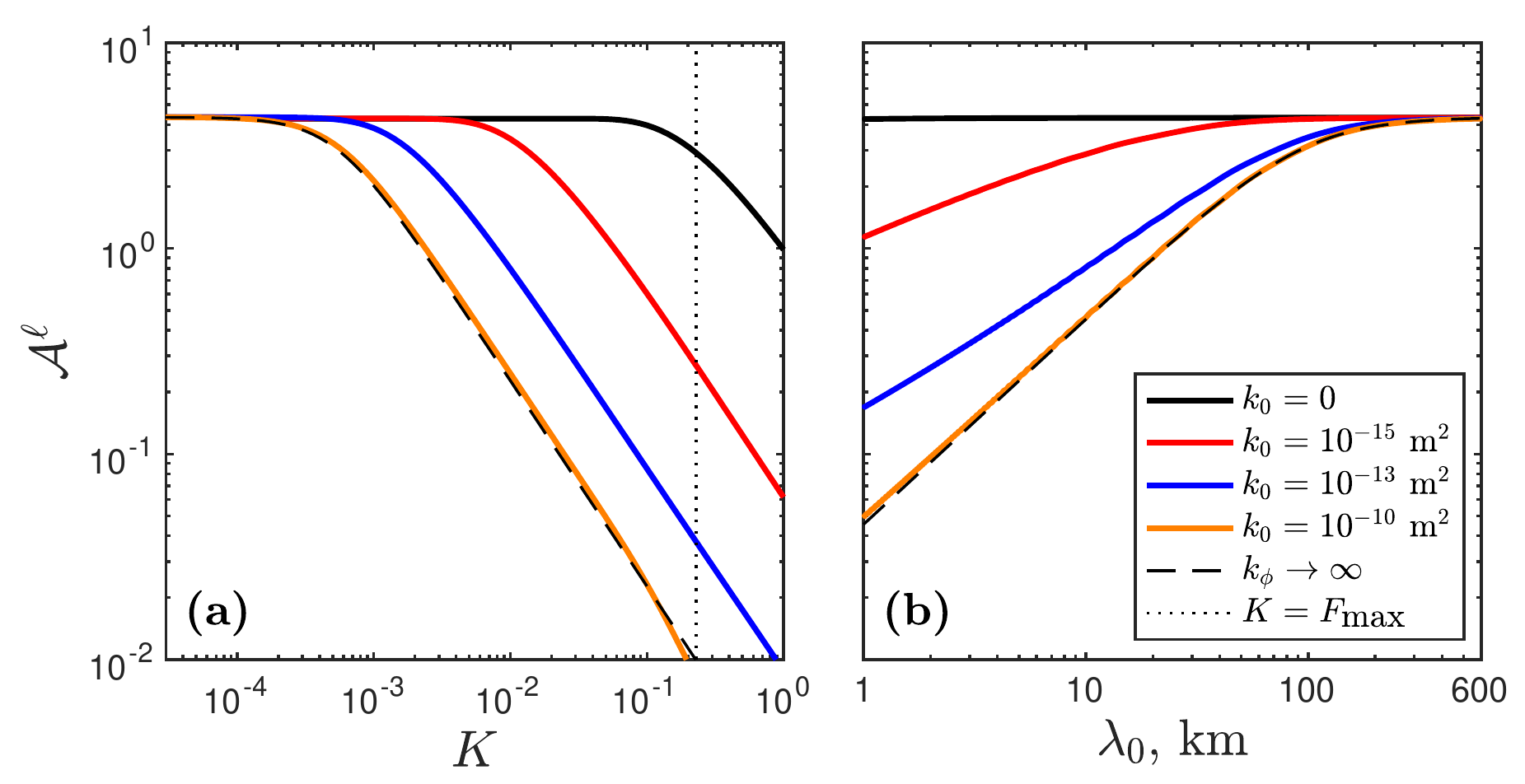}
  \caption{The control of permeability on admittance. Solid lines in
    both panels correspond to different values of reference
    permeability $k_0$. Other parameters as in
    Fig.~\ref{fig:trace_columnview}. The asymptotic solution (dashed
    line) is computed with
    equation~\eqref{eq:infinite-k-column-top}. \textbf{(a)} Admittance
    as a function of partition coefficient in a dry column for a
    heterogeneity wavelength $\lambda_0=1$~km. \textbf{(b)} Admittance
    as a function of wavelength of heterogeneity for partition
    coefficient $\pco=0.05$. The reference permeability used elsewhere
    in this paper is $k_0=10^{-12}$~m$^2$. }
  \label{fig:finite-k-admittance}
\end{figure}

Figure~\ref{fig:finite-k-admittance} shows how the rate of melt
segregation controls the admittance. If the permeability approaches
zero, solid and melt travel together and there is no phase-angle
difference: $\Delta t(F)\sim 0$. In this case, attenuation of
fluctuations is identical to dilution of the mean (this is the upper
limit of the bound~\eqref{eq:bound_liquid_admittance}).  If, at the
other extreme, the melt moves infinitely fast, then
$\Delta t(F) \sim t^s$. In this case, the liquid aggregates
instantaneous melts from the solid at all phase angles that fit
between the bottom of the column and height $Z(F)$.  For finite values
of permeability, between these two extremes, the admittance curves
take intermediate values.  As the reference permeability $k_0$ becomes
large, admittance curves in Fig.~\ref{fig:finite-k-admittance}
approach the lower-bound asymptotic result for infinite
permeability~\eqref{eq:infinite-k-column-top}.

Panel (a) of Figure~\ref{fig:finite-k-admittance} plots admittance as
a function of partition coefficient for $\lambda_0=1$~km.  Larger
partition coefficients have a taller transfer regime, providing a
longer ``runway'' for melt segregation, and hence generate phase-angle
differences that cause attenuation. Panel~(b) plots admittance as a
function of wavelength for $\pco=0.05$.  The height of the transfer
regime is fixed but as $\lambda_0$ increases, the number of
heterogeneity wavelengths that fit into the transfer regime
decreases. This reduces the phase-angle difference created by
melt segregation. 

Returning to the expression~\eqref{eq:attenuation_schematic}, we
emphasise that the dominant contribution to the admittance is
made when $F\lesssim \pco$ (when $\exp(-F/\pco)$ is of order
unity). Hence for highly incompatible elements
($\pco\ll F_\text{max}$), the ratio of wavelength to transfer-regime
height $\lambda_0/\lambda_T$ is the crucial control.  This is
expressed in equations~\eqref{eq:infinite-k-admittance} and
\eqref{eq:infinite-k-column-top}, above. In summary, the
expression~\eqref{eq:attenuation_schematic} therefore tells us that
heterogeneity wavelength, partition coefficient, adiabatic
productivity, and the rate of melt segregation are all controls on the
attenuation of trace-element variability.

With this in mind, we return to the enhanced attenuation seen in wet
melting column. There, the low-productivity tail creates a larger
$\lambda_T$ at any given value of $\pco$, as shown in
Figure~\ref{fig:Transfer_Regime}(b). Larger $\lambda_T$ allows for
more magma segregation within the transfer regime and thus greater
$\Delta t(\pco)$ and more attenuation. The comparison between wet and
dry admittance in Figure~\ref{fig:admittance_dry_wet} shows that the
ratio $\admitl_\text{wet}/\admitl_\text{dry}$ goes to $1$ when
$\pco>\Fmax$. In this range of $\pco$, $\lambda_T$ is equal to the
full column height; the effect of increasing $\lambda_T$ with a
low-productivity tail is negligible, especially since segregation is
relatively slow at small porosity.

We can also now understand the waviness of attenuation contours in
figures~\ref{fig:admittance_dry}(b) and
\ref{fig:admittance_dry_wet}(b). These oscillations appear when the
column height is similar to or greater than the height of the transfer
regime (or, equivalently, when $F_\text{max}\gtrsim D$). In these
cases, the solid throughout the column retains some of the trace
element and hence contributes to attenuation.  Then the attenuation is
higher (and $\admitl$ lower) when an integer number of solid
heterogeneity wavelengths fit into the column height. If an extra
half-wavelength fits, then $\admitl$ is higher.  For the infinite
permeability model of equation~\eqref{eq:infinite-k-admittance}, this
is expressed by the cosine term in the numerator, taking
$Z = Z_\text{max}$ for the column top.

\subsection{The role of exchange reactions toward equilibrium}
\label{sec:reaction}

In this section, we consider the exchange of trace-element mass
between solid and liquid phases that drives the system toward
equilibrium. This corresponds to the parameter regime with
$\rr>0$. The reaction rate $\rr$ is scaled by a reference melting
rate,
$\Gamma_0\equiv\rho W_0\Pi \approx 1.25\times10^{-11}$~kg/m$^3$/s,
where we used parameter values as in
Fig.~\ref{fig:dry_wet_melting_col}.

\begin{figure}[ht]
  \centering
  \includegraphics[width=\textwidth]{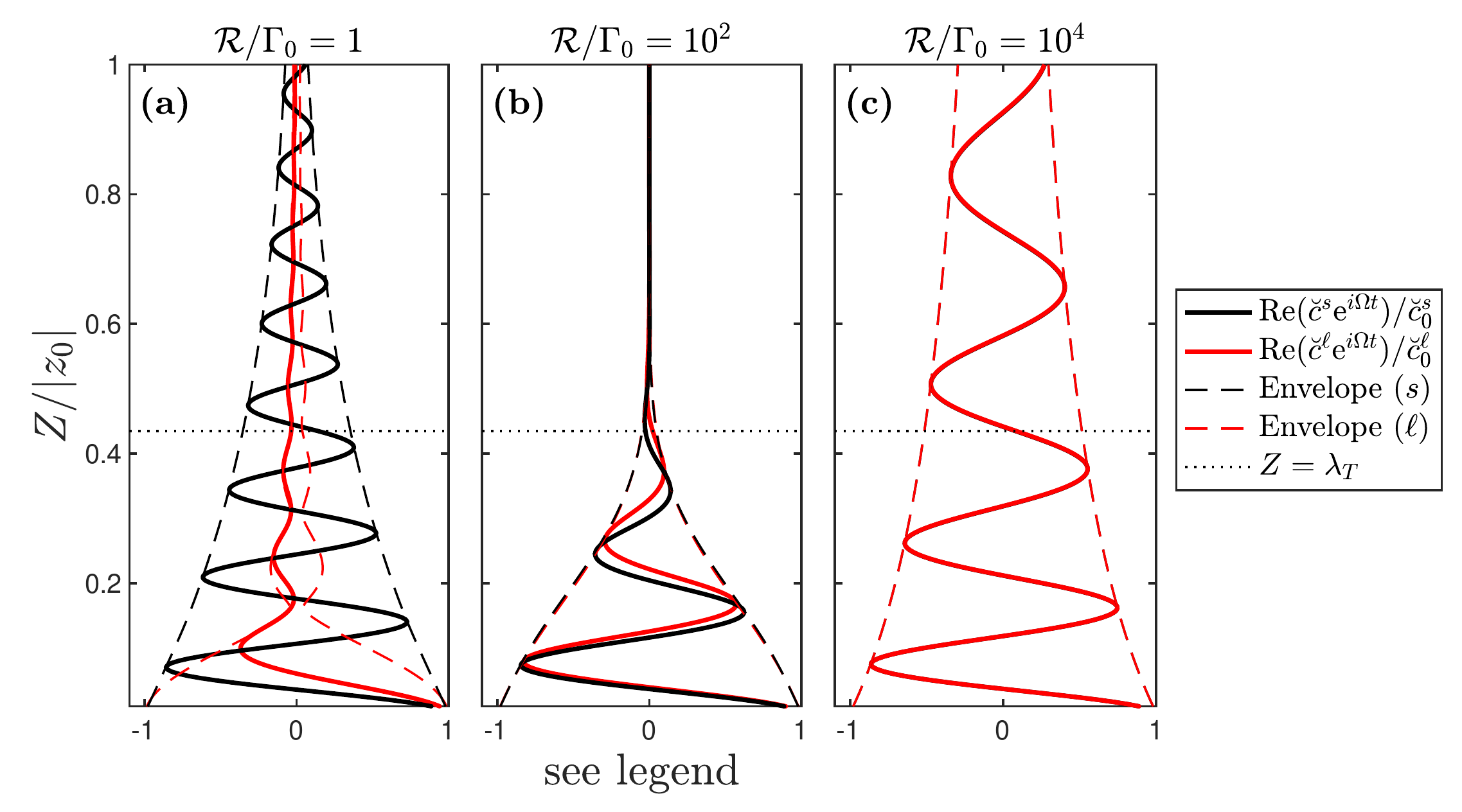}
  \caption{Vertical structure of fluctuations in the liquid and solid
    phase, $\real(\csf \text{e}^{i\freq t_0})$ (black) and
    $\real(\clf \text{e}^{i\freq t_0})$ (red), of a trace element with
    $\pco=0.1$ and $\lambda_0=10$~km, for three different reaction
    rates.  Melting is computed assuming a dry column. Solid
    fluctuations are normalised by the initial value in the unmelted
    mantle source; liquid fluctuations are normalised by that in the
    incipient melt.  The scaled reaction rate is \textbf{(a)}
    $\rr/\Gamma_0=10^{-3}$, \textbf{(b)} $\rr/\Gamma_0=5$ and
    \textbf{(c)} $\rr/\Gamma_0=10^3$.  The solid lines are plotted for
    an arbitrarily chosen time $t$; dashed lines show the envelope of
    fluctuations.}
  \label{fig:trace_columnview_react}
\end{figure}

Figure~\ref{fig:trace_columnview_react} shows trace-element
concentration in the liquid and solid for three values of $\rr$ that
span the behavioural spectrum. The column has dry melting with
$\pco=0.1$ and $\lambda_0=10$~km.  In panel~(a), $\rr=0$ (as in the
sections above), giving complete disequilibrium transport; the
phase-angle difference between the liquid and solid phases in the
transfer regime controls the attenuation.  In panel~(c), the reaction
rate is large enough that the trace element is in approximate
equilibrium: $\csf \approx \pco \clf$ for all $Z$.  The liquid and
solid fluctuations remain in phase throughout the column and move
upward with the chromatographic velocity \citep{navon87}. Attenuation
in this quasi-equilibrium case is independent of $\lambda_0$; instead
it depends only on $\pco/\Fmax$. Indeed, below we demonstrate that
admittance is generally maximised for $\rr\to\infty$.

Figure~\ref{fig:trace_columnview_react}(b) shows the case of
intermediate $\rr$, where exchange reactions move the system toward
trace-element equilibrium but are not fast enough to achieve it. The
phase-angle difference between the solid and liquid curves is
non-zero. Attenuation of liquid fluctuations occurs by interphase
transfer, but it also occurs by exchange reactions.  This
combination can lead to greater attenuation (and hence smaller
$\admitl$) than at either of the reaction-rate extremes.

\begin{figure}[ht]
  \centering
  \includegraphics[width=\textwidth]{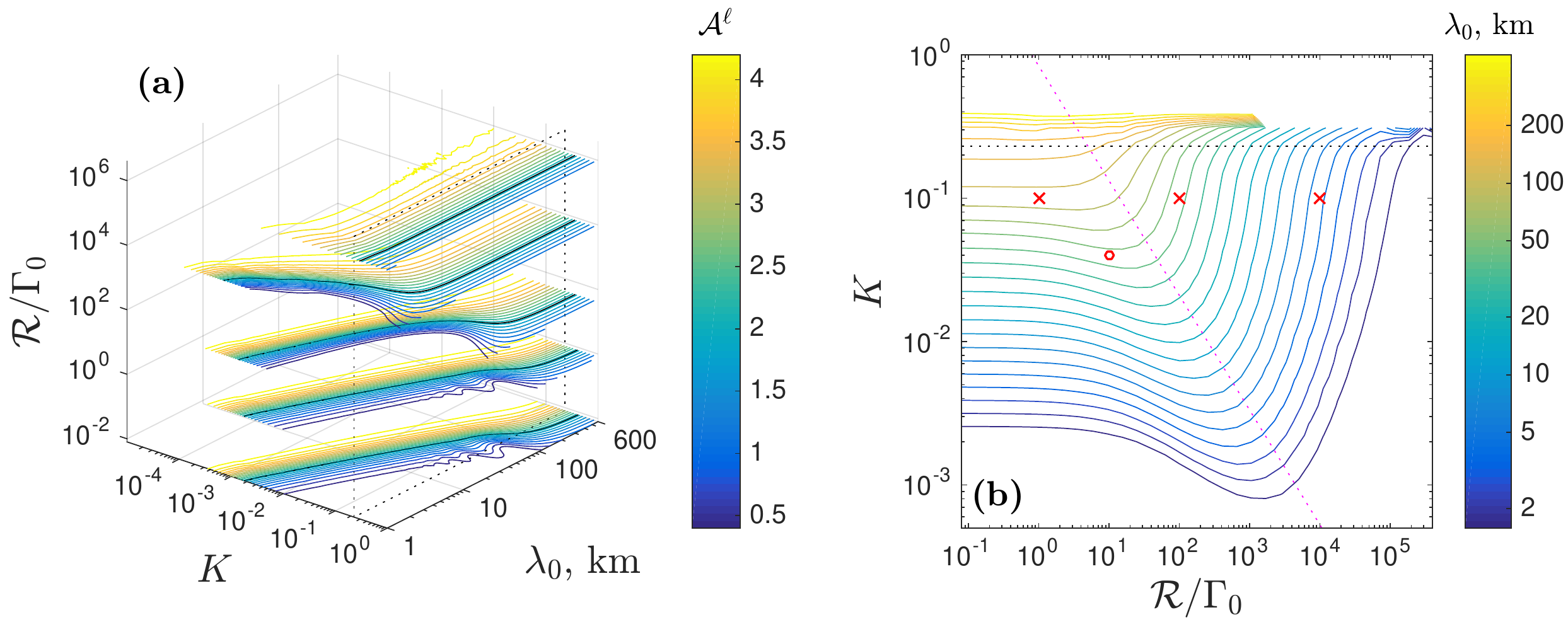}
  \caption{The systematics of liquid admittance as a function of
    partition coefficient, heterogeneity wavelength and reaction rate
    for dry melting. \textbf{(a)} Contours of $\admitl$ in the
    three-dimensional space of $\pco$, $\lambda_0$, $\rr/\Gamma_0$ at
    values of
    $\rr/\Gamma_0 = 8\times\{10^{-3}, 10^{-1}, 5, 10^3, 10^5\}$.
    Black contour shows $\admitl=2$. The set of contours at the
    smallest value of $\rr/\Gamma_0$ are almost identical to those in
    Fig.~\ref{fig:admittance_dry}(b). \textbf{(b)} Contours of
    $\lambda_0$ on a surface defined by $\admitl=2$. Points marked
    with a red $\times$ are the conditions of the three panels in
    Fig.~\ref{fig:trace_columnview_react}; the red circle corresponds
    to $\pco=0.04,\rr/\Gamma_0=10$. The magenta dotted line has a
    slope of~-1. In both panels, the black dotted lines indicate the
    position of $\pco=\Fmax$.}
  \label{fig:admittance_reactionrate}
\end{figure}

Figure~\ref{fig:admittance_reactionrate} shows the systematics of
$\admitl$ as a function of $\pco$, $\lambda_0$ and $\rr$ for dry
melting-column calculations. Panel~(a) displays the full,
three-dimensional space with contours of $\admitl$ plotted at five
values of $\rr/\Gamma_0$.  First we consider the set of contour lines at
the smallest value of $\rr/\Gamma_0$. These are nearly identical to the
contours in Fig.~\ref{fig:admittance_dry}(b) because reaction plays
almost no role in equilibrating the solid and liquid. In this set of
contours, at wavelengths $\lambda_0\gtrsim100$~km, the admittance
becomes nearly independent of $\lambda_0$ because there is almost no
phase-angle difference between the solid and the liquid concentration
profiles.  Hence for very large wavelengths of heterogeneity, the
system is in approximate equilibrium with respect to the partition
coefficient despite melt segregation and the lack of reaction.

Moving to higher reaction rates, the quasi-equilibrium regime extends
toward smaller wavelengths.  This is because reaction tends to
eliminate any phase-angle difference that would be created by
segregation (cf.~Fig.~\ref{fig:trace_columnview_react}c). For the
fastest reaction rates considered, admittance becomes independent of
wavelength for heterogeneities at scales greater than 1~km.

Another notable feature of Figure~\ref{fig:admittance_reactionrate}a
is evident by comparison of all sets of contours at $\pco=0.1$ and
$\lambda_0=10$~km (cf.~Fig.~\ref{fig:trace_columnview_react}a and
b). Under these conditions, admittance decreases with increasing
reaction rate and then increases again.  The former is due to reaction
acting on (but not eliminating) differences in phase angle; the latter
occurs as reaction drives the system into the equilibrium regime.

Figure~\ref{fig:admittance_reactionrate}b is a different view of the
effect of reaction rate.  Here we plot contours of the wavelength
$\lambda_0$ at which $\admitl=2$. The contours indicate the smallest
wavelength of heterogeneity that can be preserved under various
conditions of reaction rate and partition coefficient. Following a
horizontal line at, say, $\pco=10^{-2}$ from low to high $\rr$,
wavelength increases slightly (more attenuation due to reaction)
before decreasing sharply (less attenuation in the quasi-equilibrium
regime).  The sharp change from the disequilibrium regime to the
quasi-equilibrium regime occurs across a boundary with a slope of $-1$
on this diagram.

Experimental measurements of trace element diffusivity indicate that
it is extremely small \citep{vanorman01}.  For example, for Neodymium
in a spherical grain of radius $a=3$~mm at a pressure of 1~GPa and
temperature of 1300$^\circ$C, the reaction rate would be
\begin{equation}
  \label{eq:reaction_rate_estimate}
  \rr \sim \frac{4\pi\rho\diffusivity}{a^2} \approx 1\times 
  10^{-10}\text{ kg/m$^3$/s},
\end{equation}
where $\diffusivity$ is the diffusivity in the solid. This estimate
corresponds to $\rr/\Gamma_0\approx 10$.  For a partition coefficient
of $\pco\approx0.04$, this sits in the disequilibrium regime (red
circle in Fig.~\ref{fig:admittance_reactionrate}(b)), but is rather
close to the transition to chromatographic transport.

Cast in terms of a characteristic equilibration time, the above gives
approximately one million years for Nd.  At intermediate mid-ocean
ridge spreading rates, one million years is enough time for solid
mantle to upwell through roughly half of the silicate melting regime
beneath the axis.  Hence, for $\sim$3 millimetre grain size, we
consider diffusive reequilibration of trace elements to be slow. But
the quadratic dependence of $\rr$ on grain size means that smaller
grains will equilibrate much faster.  There are few constraints on
grain size in the asthenosphere, however, and models remain
speculative.

Differences in diffusivity between trace elements may help to explain
anomalies in their behaviour, relative to a model based on equilibrium
partitioning.  These effects would be of second order, however,
whereas the questions motivating this study pertain to observations of
first-order patterns.

\section{Comparison with observations}
\label{sec:observations}

Model predictions can be compared with observations of trace-element
variability by making assumptions about the characteristics of
heterogeneity that enters the bottom of the melting column. In
particular, we must prescribe a time-series of concentration for each
trace element in the source mantle.  This is largely unconstrained and
so we make simplifying assumptions. The key assumption is that the
input heterogeneity is identical for all trace elements, i.e., it is
independent of $\pco$.  The theoretical framework proposed here
requires only that the time-series be periodic; we can then analyse it
in terms of its decomposition into Fourier modes. Below, after a
discussion of the geochemical datasets, we formulate a synthetic
representation of periodic heterogeneity that is suitable.

In section \ref{sec:model-data}, we discuss the synthetic
heterogeneity signal and describe models that aim to fit observational
data. We use only dry column models but consider mantle heterogeneity
with different periodicity, for comparison with observations. Then, in
section~\ref{sec:observation-data}, we summarise published geochemical
observations from eruptions in Iceland and from a set of MORBs sampled
from the Central Indian Ridge. The data are considered in terms of
their variance for each measured trace element. Importantly, the
datasets all show a roughly log-normal distribution of concentrations
for each element.  This motivates a hypothesis for the form of a
synthetic heterogeneity.

\subsection{Synthetic heterogeneity}
\label{sec:model-data}

Constructing model instances to compare with observations involves
specifying the parameters of the melting column (e.g., $z_0$,
$F_\text{max}$) as well as the details of the input heterogeneity.
Thus far, we have considered only heterogeneity patterns consisting of
sinusoids of a single frequency.  But the theory is linear and hence
superpositions of such sinusoids are also valid solutions. This opens
a very large parameter space.  For example, one could consider all
heterogeneity signals that are formed by assigning a linear slope
$\beta$ to the power-spectral density within the wavelength band
associated with mantle heterogeneity \citep[e.g., a white
spectrum,][]{gurnis88}.

\begin{figure}[ht]
  \centering
  \includegraphics[width=0.7\textwidth]{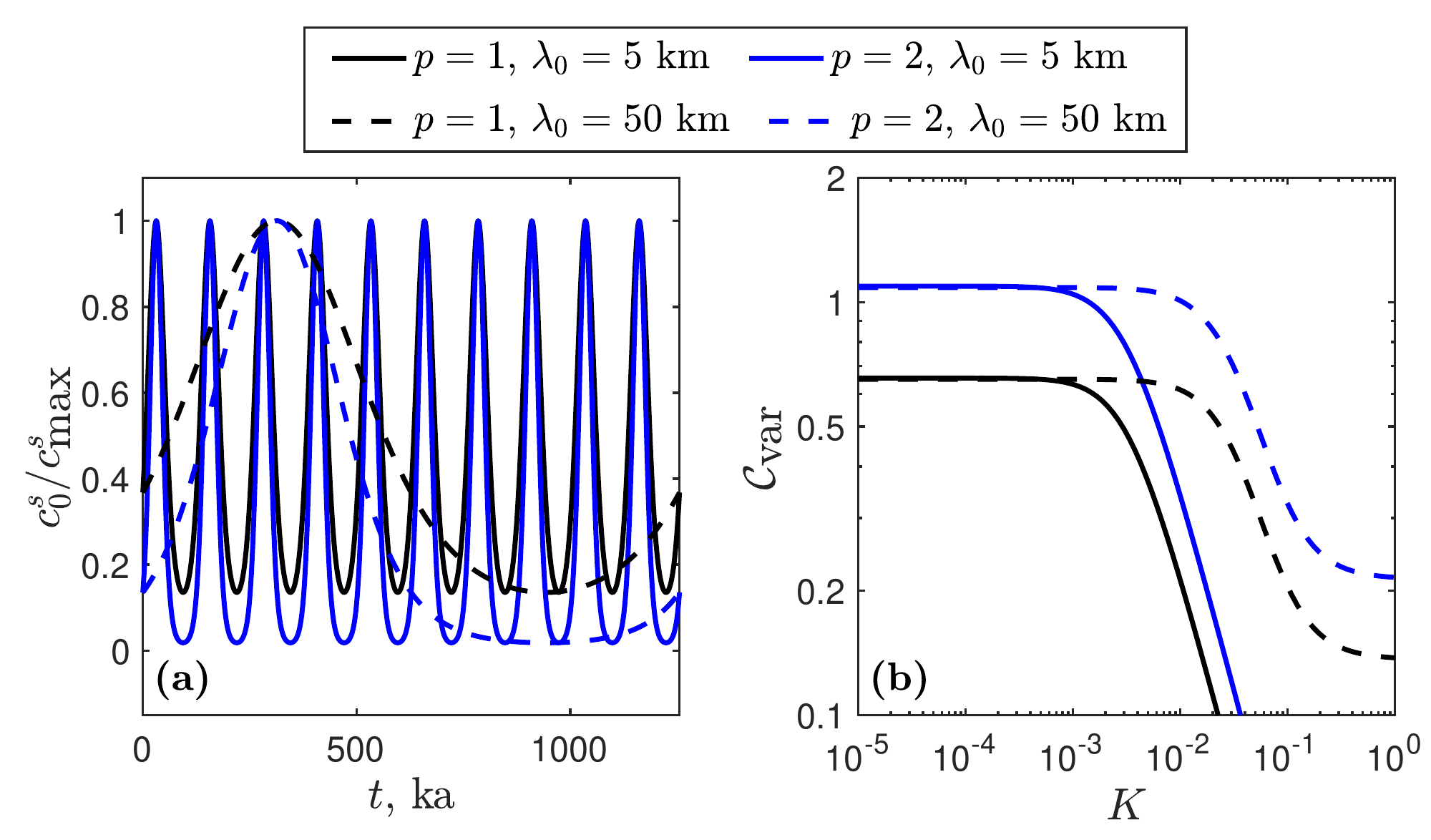}
  \caption{The log-sinusoidal heterogeneity signal for $\sha=1, 2$ and
    $\lambda_0=5, 50$~km. \textbf{(a)} The heterogeneity signal as a
    function of time. \textbf{(b)} The coefficient of variation at the
    top of the column as a function of partition coefficient. The dry
    column uses $W_0=4$~cm/yr, $\rr/\Gamma_0=0$ and other parameters
    as in Fig.~\ref{fig:trace_columnview}.}
  \label{fig:log_heterogeneity}
\end{figure}

For present purposes, we adopt a simpler approach: we choose a
periodic function that can be tuned to give a suitable maximum
variance. Hence it is sufficient for comparison with the data
distribution but without additional, unconstrained complexity.  In
particular, we propose the following log-sinusoidal form for the
source heterogeneity,
\begin{equation}
  \label{eq:log-sinusoidal-function-bottom}
    \cs_0(t) = \cs_\text{max}\e^{\sha\left(\sin \freq t -1\right)},
\end{equation}
for $\sha>0$ where $\cs_\text{max}$ is the maximum concentration (which
does not need to be specified).  This function is plotted for two
values of $\sha$ and two values of $\freq_0 = 2\pi W_0/\lambda_0$ in
Figure~\ref{fig:log_heterogeneity}(a). It is similar in form to the
Gaussian pulse-train proposed by \cite{liang18}.

Since the geochemical column models developed above are based on a
time-dependence expressed by $\e^{i\freq t}$, we express the synthetic
heterogeneity function~\eqref{eq:log-sinusoidal-function-bottom} in
terms of the coefficients of a Fourier series
\begin{equation}
  \label{eq:log-sinusoidal-Fourier-bottom}
  \cs_0(z=z_0,t) = \csm_0 + \sum_{j=1}^\infty(a_j\cos j\freq_0 t 
  + b_j\sin j\freq_0 t).
\end{equation}
Coefficients $a_j$ and $b_j$ are determined numerically.

The liquid concentrations at the column top can also be expressed as a
Fourier expansion, but with different coefficients, $a'_j,b'_j$.
Because the column model is linear, the primed Fourier coefficients
are related to unprimed coefficients by
\begin{equation}
  \label{eq:log-sinusoidal-Fourier-top}
  a'_j + i b'_j =
  \admitl(\lambda_0\vert\pco,\rr)\e^{i\Delta\theta_j}\times(a_j + i b_j),
\end{equation}
where $\Delta\theta_j$ is the phase-angle difference between the
column bottom and top for each mode.  The primed coefficients and the
column-top mean liquid concentration are used to invert the Fourier
series for the concentration time-series at the top of the column.

\subsection{Geochemical data}
\label{sec:observation-data}

We consider measurements of trace-element concentrations in
mantle-derived basalts from three datasets that, in broad terms,
represent three different timescales of magma genesis, segregation and
eruption.

The first, termed the ``Iceland Single Eruption'' contains
olivine-hosted melt-inclusion data from the Haleyjabunga eruption of
southern Iceland \citep{neave18}.  Melt inclusions may capture more
mantle-derived variability in melt chemistry compared with their
associated whole rock, because they are trapped before extensive
crustal mixing has occurred \citep[e.g.,][]{sobolev93,
  sobolev1996_petrology, maclennan08}.  Iceland's geology provides a
unique constraint on magma residence time in its crust: glacial
unloading at the end of the last ice age generated enhanced melting in
the shallow melting region, supplying a burst of
incompatible-element-depleted melts \citep{jull96}.  These melts
erupted within 1000~years of deglaciation occurring
\citep{maclennan02}, which provides the upper bound on the
source-to-surface magma transport and residence time beneath Iceland.
This timescale is effectively instantaneous in terms of solid mantle
upwelling.

The second dataset, termed ``Iceland Multiple Eruptions,'' uses the
compilation from \cite{shorttle11} and includes whole-rock data from
Iceland's northern neovolcanic zone. These glacial and post-glacial
eruptions represent a medium timescale of mantle sampling of probably
less than $100$~kyr.

The third dataset, termed ``MORB Series,'' comes from
\cite{cordier10}, who analysed samples from the Central Indian Ridge,
which spreads at a full rate of 42~mm per year
\citep{demets90}. Off-axis samples, collected by submersible, extend
their record back $\sim800$~kyr. They document a chemical periodicity
that is symmetric across the ridge axis at a period of
150--200~ka. Multiplying by an appropriate corner-flow upwelling
speed, this periodicity would correspond to mantle heterogeneity at a
wavelength of order 10~km.

\begin{figure}[ht]
  \centering
  \includegraphics[width=\textwidth]{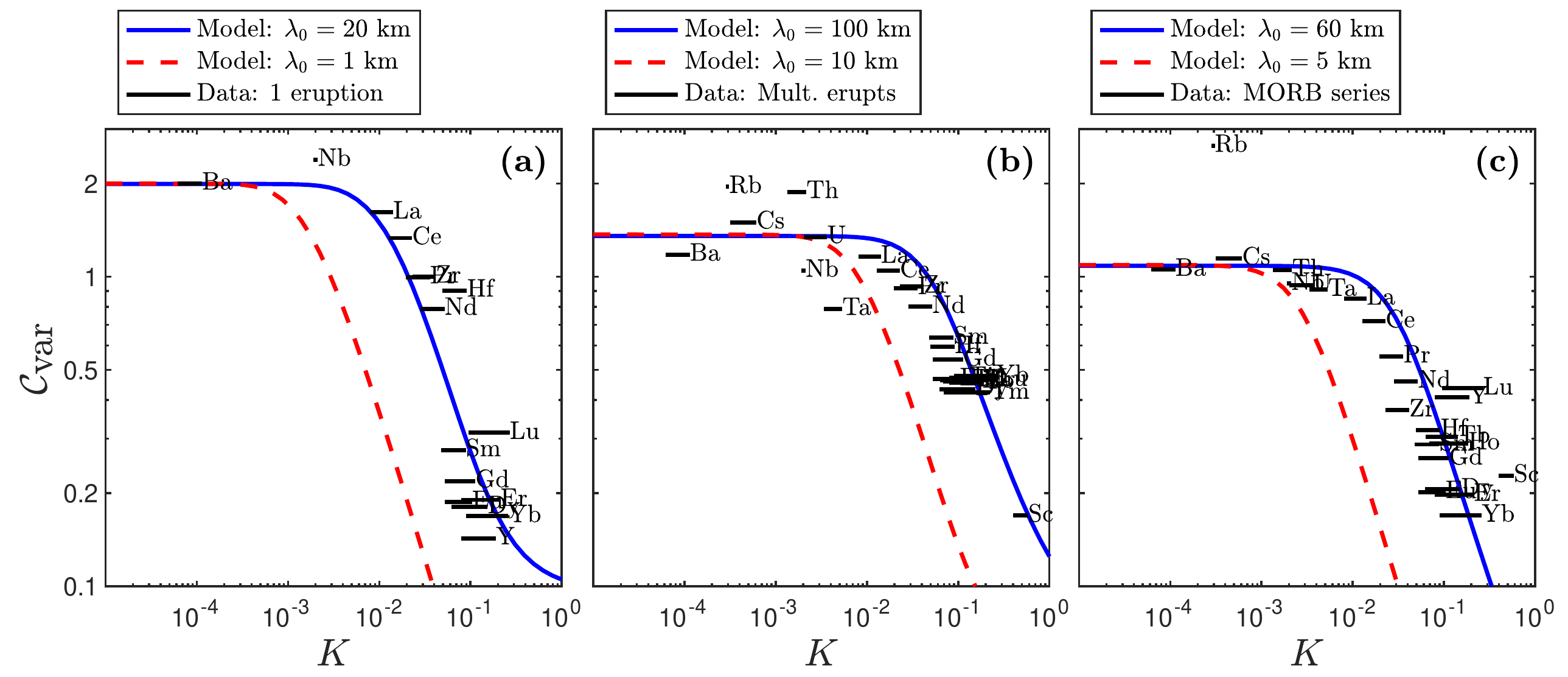}
  \caption{Coefficient of variation $\cvar$ for various trace elements
    in three different datasets. Model curves are overlayed.  Solid
    blue curves represent numerical results fitted to data by
    adjusting the heterogeneity wavelength only.  Dashed red curves
    represent numerical results with a closest timescale to
    geochemical data. Black lines mark $\cvar$ of geochemical data
    with a range of $\pco$.  \textbf{(a)} ``Iceland Single Eruption''
    \citep{neave18}; \textbf{(b)} ``Iceland Multiple Eruption"
    \citep{shorttle11}; \textbf{(c)} ``MORB Series''
    \citep{cordier10}. An upwelling rate $W_0$ of 1~m/yr is assumed
    for the Iceland models whereas a value of 2.8~cm/yr is chosen for
    the Central Indian Ridge.  Partition coefficients are from
    \citet{neave18}, with the width of the bar representing the range
    in partition coefficient between spinel- and garnet-field
    melting.}
  \label{fig:comparison-with-data}
\end{figure}

Data are plotted in Figure~\ref{fig:comparison-with-data}, with the
three datasets shown separately in panels (a)--(c).  For any trace
element, the samples in each dataset are distributed roughly according
to a log-normal distribution.  The distribution for each element is
summarized in terms of the coefficient of variation $\cvar$,
\begin{equation}
  \label{eq:define_cvar}
  \cvar = \sigma/\mu,
\end{equation}
where $\sigma$ is the standard deviation of the concentrations and
$\mu$ is the mean. This formula is applied to the data and the models.
In Fig.~\ref{fig:comparison-with-data}, $\cvar$ is plotted as a
function of the bulk partition coefficient. For each trace element in
the data, $\pco$ is estimated using a peridotitic mineralogy.  The
uncertainty in $\pco$ represents the difference between partitioning
in the garnet and spinel stability fields (garnet generally gives
a higher $\pco$).

The data in Fig.~\ref{fig:comparison-with-data} show an obvious trend
with partition coefficient.  At small $\pco$, the coefficient of
variability is large --- between one and two times the mean. There is
some scatter in $\cvar$, but it generally shows a plateau for
$\pco\lesssim 10^{-2}$; at higher values of $\pco$, $\cvar$ declines
sharply.

Modelling results are compared with data in Figure
\ref{fig:comparison-with-data}. Solid blue curves represent the
best-fitting numerical results for each dataset, while red dashed
curves are numerical results with a wavelength that is closest to the
geochemical timescale of the data.  Panel (a) shows the comparison
with ``Iceland Single Eruption;" $\sha=8$ in this model. The
best-fitting curve has a heterogeneity wavelength of $20$~km and shows
a good fit to the data, reasonably matching all elements except for
niobium. However, the timescale of chemical variation at the column
top that is associated with $\lambda_0=20$~km is
$\tau = \lambda_0/2W_0 \approx 20$~kyr (assuming an upwelling rate of
$1$~m/year).  It is very unlikely that melt inclusions from 20~kyr of
magmatic accumulation would appear in the same eruption.  A more
realistic period of accumulation is less than $1$~kyr, represented by
the red dashed curve.  However, this timescale corresponds to a
smaller wavelength of heterogeneity that is more attenuated than
observed.  This red curve could be shifted to larger admittance by
assuming a smaller permeability $k_0$ (as in
Fig.~\ref{fig:finite-k-admittance}).  However, to shift upward by a
factor of $\gtrsim 5$, as required to fit the data, would mean
decreasing $k_0=10^{-12}$~m$^2$ by between two and three orders of
magnitude. We recall that $k_0$ is the permeability at reference
porosity $\phi_0=1$\%, and that the reference speed of melt
segregation is then $w_0 = k_0\Delta\rho g / \phi_0\mu$. Hence
$k_0=10^{-14}$~m$^2$ corresponds to a speed of about 2~mm per year,
which is inconsistent with constraints from uranium-series
disequilibrium \citep{stracke06}.

Panel (b) of Fig.~\ref{fig:comparison-with-data} compares the
``Iceland Multiple Eruption'' dataset with models using $\sha=3$. The
best-fitting curve, with a wavelength of $100$~km and a timescale of
$100$~kyr, can fit most of the data. This wavelength also corresponds
to a reasonable geochemical timescale. The model curve based on a
wavelength of $\lambda_0=10$~km provides a poor fit to the data but
has an acceptable geochemical timescale ($\sim10$~kyr).

Data/model comparison with ``MORB series'' in panel~(c) is consistent
with the comparison for Icelandic basalts.  The geochemical periodicity
($\sim$175~ka) that was identified by \cite{cordier10} is associated
with a $\sim$5~km wavelength of heterogeneity (assuming an upwelling
rate of 2.8~cm/year), whereas the best fitting wavelength of 60~km
would have a periodicity of $2.1$~Myr, longer than the timescale
sampled by the entire dataset \citep[800~kyr,][]{cordier10}.  A value
of $\sha=2$ is used in this case.

Curves in Figure~\ref{fig:comparison-with-data} are computed with a
dry column model with constant isentropic productivity.  If we had
instead used the wet model, the admittance at all but the smallest
partition coefficients would be reduced.  To compensate for this, a
larger wavelength would be needed to fit the observations.  This would
put the model even further outside the timescale constraints
associated with the data.

\section{Discussion}
\label{sec:discussion}

In this section we discuss aspects of the results above, in comparison
with observations and with other relevant constraints.  We summarise
the systematics of the model and highlight its deficiencies (in the
narrow sense of the approximations made).  We then discuss the model
in the broader context of models that could plausibly explain the
observations, including the other end-member explanation of
heterogeneity of melt-transport processes.  We conclude with some
remarks on the path forward.

\subsection{A correct and sufficient explanation of the observations?}

Evidently, the column models (and synthetic heterogeneity) developed
here can provide a good fit to the variability spectrum of trace
elements in several natural settings.  This is because the models and
data share two key characteristics. First, a plateau in the
coefficient of variability at the smallest partition coefficients.
And second, a sharp drop-off in variability with increasing partition
coefficient.  The model is matched to these characteristics by
adjusting the $\sha$ value of the synthetic heterogeneity, which controls
the sharpness of the enriched peaks, and its fundamental wavelength
$\lambda_0$.  The former sets the height of the plateau in $\cvar$ at
small $\pco$ whereas the latter controls the position of the drop-off
in $\cvar$ at larger $\pco$.

Does the goodness of fit between models and data, then, indicate that
the models are a correct and sufficient explanation for the
observations? Almost certainly not.  The synthetic heterogeneity used
here is undoubtedly oversimplified from the natural system, but we
have few constraints on what it really should be.  Moreover, since we
consider only variability for each trace element, there are other
synthetic patterns that would have worked equally well (for example,
the family with the same power spectrum but with randomised phase
angles). The more significant problem is the fundamental wavelength,
$\lambda_0$.  

The best-fitting wavelengths in Figure~\ref{fig:comparison-with-data}
are relatively large, which gives rise to an important discrepancy
with observations. Consider, first, the single eruption in panel~(a).
For a best-fitting wavelength of 20~km and an upwelling speed of
1~m/year, the period of chemical oscillation in our column model would
be 20~ka.  In contrast, the melt-extraction time-scale in Iceland is
probably on the order of thousands of years.  The magma that was
captured in the melt inclusions of the single eruption analysed by
\cite{neave18} was probably generated over a period similar to the
melt-transport time-scale --- a factor of twenty smaller than
suggested by the model. A heterogeneity wavelength that is consistent
with the melt-transport time-scale, $\lambda_0=1$~km, gives a model
outcome that is inconsistent with observations.

The MORB series from the Central Indian Ridge
(Fig.~\ref{fig:comparison-with-data}(c)) presents a similar issue.
The best fitting wavelength corresponds to a period of just over two
million years (assuming upwelling at 2.8~cm/year). But the time-span
of the observations, judging from the spreading rate and the off-axis
distance, is about one million years \citep{cordier10}. Moreover,
there appear to be about five geochemical ``cycles'' within this
period, rather than the half-cycle that would be predicted for
$\lambda_0=60$~km.  So again, the time-period associated with the
best-fitting wavelength represents a discrepancy with
observations. Taking a wavelength of 5~km to roughly match the period
of the observed geochemical cycle leads to a model $\cvar$ curve that
is inconsistent with the data.

The same issues applies in comparison between the model and the
Iceland Multiple Eruptions series
(Fig.~\ref{fig:comparison-with-data}(b)), though it is less severe.
The time-span of the eruptions is $\sim100$~ka, which is the same as
the period of the best-fitting oscillation (for upwelling at
1~m/year). This means that a single cycle of heterogeneity has passed
through the system during the recorded eruptions.  The data, however,
show no evidence for the systematic temporal variation that might be
expected with this period \citep{shorttle11}.  A heterogeneity
wavelength of 10~km, also plotted in panel~(b), provides a poor fit to
the data. However, it gives an indication of the model sensitivity to
wavelength: the curve denoting $\cvar$ shifts to smaller $\pco$ by one
order of magnitude, which is as predicted by our asymptotic model
(eqn.~\eqref{eq:infinite-k-column-top}).

It is unlikely that the contribution of off-axis melting would resolve
this discrepancy.  Lateral focusing of magma \citep[e.g.,][]{sparks91}
brings the output of off-axis columns to the ridge axis, where it
presumably mixes with the melt produced directly below the
ridge. Off-axis columns are shorter and melt to lower
$\Fmax$. However, at moderate distances off axis and for most
incompatible elements we have $\Fmax/\pco \gg 1$, and so the
admittance spectrum should be similar on- and off-axis. More
importantly, if the pattern of heterogeneity in the mantle is
isotropic (i.e., equant heterogeneities), then we expect incoherence
of phase-angle between on- and off-axis columns. Indeed, mantle
heterogeneities would need to be elongate and roughly sub-parallel to
the base of the lithosphere for their signal to sum coherently at the
ridge axis.  There is no reason to expect this to be the case; indeed
\textit{a priori}, incoherence and cancellation is the most likely
scenario. This is especially true at wavelengths smaller than the
maximum lateral focusing distance. Accounting for melt from off-axis
columns would thus increase the discrepancy with the observed
time-scale.

Therefore, while the good correspondence between models and
observations in Figure~\ref{fig:comparison-with-data} is intriguing,
it cannot be interpreted as a validation of the model. The end-member
of filtration of trace-element heterogeneity by vertical migration and
aggregation of fractional melts is not a sufficient explanation for
the observations. Despite this, the comparison does not exclude the
possibility that such filtration contributes to observed patterns.
Indeed it may be possible to discern its effects in more elaborated
models such as those discussed below.

\subsection{Model systematics and limitations}

We here summarise and critique the model proposed above.

Our definition of admittance means that the filtration
properties of the melting column are captured by its systematics.
This is best summarised by the asymptotic solution for
infinite permeability (eqn.~\eqref{eq:infinite-k-column-top}). It
shows that attenuation of amplitude for a particular mode is expected
when the wavelength of that mode is small compared to the height of
the transfer regime.  A smaller $\pco$ means a shorter transfer regime
and hence less attenuation of heterogeneity at a given wavelength.
Small amounts of reactive equilibration enhance the attenuation of
heterogeneity.  It is only at the highest reaction rates (e.g., for
grain sizes of tens of microns) that near-chromatographic transport
occurs, preserving heterogeneity at all wavelengths.

The asymptotic solution assumes that the isentropic productivity is
uniform with depth.  This is a reasonable approximation for a dry
melting column, but not when volatiles are present.  In that case, a
low-productivity zone appears at the base of the melting region and
lengthens the transfer regime.  Porosity, permeability and hence melt
segregation are small in this zone compared with the silicate melting
region above.  Nonetheless, segregation over the longer transfer
regime reduces the admittance for most $\pco$ values. The alternative
scenario to this is one where productivity is high at the base of the
melting region, such as occurs for some pyroxenitic lithologies
\cite[e.g.,][]{lambart2016_jgr}.  In this case, the transfer regime would be
diminished in height; melt segregation would be enhanced by higher
porosity, but the overall effect would be to increase admittance for
most $\pco$ values.  This highlights the importance of melt
productivity at the onset of melting for attenuation of source
heterogeneity.

The present model is clearly an end-member of the possible models for
trace-element variability in basalts. Below we discuss it within this
broader context. However, even in the narrow confines of
one-dimensional column models, there is an assumption made above that
should be questioned. We have postulated fractional melt production
$(\cg=\cs/\pco)$ while also requiring negligible reactive
equilibration $(\rr=0)$.  However, production of incremental melts
that are in equilibrium with the solid concentration requires that
trace-element mass is rapidly transferred from the interior of solid
grains to their rim, which contradicts the choice of $\rr=0$. A
treatment of incremental melts consistent with $\rr=0$ is $\cg=\cs$,
where $\cs$ is the mean concentration of the solid. If the solid
grains are initially uniform in concentration then, with this
combination of $\cg$ and $\rr$, they remain uniform; hence the
concentration at the rim of the grain is equal to the mean
concentration.  But this concentration can be far from obeying the
partitioning behaviour that is observed in laboratory experiments and
natural lavas; it is therefore dismissed on empirical grounds.

Previous workers have proposed models that reconcile these
contradictions. This class of models resolve the solid concentration
as a function of radius within the interior of representative grains
\citep[e.g.,][]{qin92, iwamori93a, liang03}. Chemical diffusion in the
radial direction allows for transport of trace elements to the rim of
the grain, where they are transferred to the melt according to their
concentration (and concentration gradient) at the rim. This approach
should be applied to the problem of sinusoidal variation of trace
elements in the source. While that is beyond the present scope, it is
worth considering the time-scales associated with the relevant
processes: intra-grain diffusion of concentration, melting from $F=0$
to $F=\pco$, and variation of concentration by melt segregation from a
heterogeneous source.  These can be written
\begin{subequations}
  \label{eq:timescales}
  \begin{align}
    \tau_\text{difn} &\sim \frac{a^2}{4\pi\diffusivity}\approx 200\text{ ka},\\
    \tau_\text{melt} &\sim \frac{\pco z_0}{\Fmax W_0}\approx 800\text{ ka},\\
    \tau_\text{hetr} &\sim \frac{\lambda_0}{W_0}\approx 100\text{ ka},
  \end{align}
\end{subequations}
where we have used grain size $a=3$~mm, diffusivity
$\diffusivity=10^{-19}$~m$^2$/s, partition coefficient $\pco=0.01$,
column height $z_0=70$~km, maximum degree of melting $\Fmax=0.23$,
upwelling speed $W_0=4$~cm/y and source heterogeneity wavelength
$\lambda_0=5$~km. The ratio
$\tau_\text{difn}/\tau_\text{melt}\approx1/4$ tells us that diffusion
is moderately faster than melting.  It is independent of the
wavelength of heterogeneity, but is sensitive to the grain size. For
the assumption of $\cg=\cs/\pco$ to be justified, we'd need diffusion
within the grain to be much faster than melting.

The ratio $\tau_\text{difn}/\tau_\text{hetr}\approx 2$ tells us that
diffusion is commensurate with or slightly slower than fluctuations
due to heterogeneity.  This number depends on grain size and
heterogeneity wavelength. To properly justify the assumption of
$\rr=0$, the timescale of diffusion should be much greater than that
of chemical variability due to heterogeneity (so
$\tau_\text{difn}/\tau_\text{hetr}$ should be very large).  From these
arguments we can conclude that the model assumptions made here, while
effective for simplifying the problem, cannot be justified robustly on
the basis of scaling arguments.

However, geochemical observations of mean trace-element concentrations
have long been interpreted in terms of fractional melting. Therefore
this assumption is scientifically relevant.  Further work is needed to
develop a theory for admittance of trace-element heterogeneity in the
context of grain-resolving models, building on the existing literature
\citep[e.g.][]{qin92, iwamori93a, liang03}.

Other column models, going back at least to \cite{mckenzie85}, have
allowed for a parameterised lateral transport of magma into isolated
channels with rapid transport to the surface.  This approach has been
further formalised in terms of a ``double-porosity'' theory, with
overlapping and coupled continua representing the high-permeability
channels and the low-permeability inter-channel regions separately
\citep{liang10a}. Pseudo-two-dimensional models by \cite{liu17} apply
the double-porosity theory to isotope systems beneath a mid-ocean
ridge.  Models with one porosity field that resolve the dynamics in
2-D show that channelised transport can generate chemical variability
from a homogeneous mantle \citep{spiegelman03a}. However,
\cite{liang11} cautioned that porosity waves associated with reactive
flow can promote dispersion and mixing of chemical
heterogeneities. \cite{liang18} found that an isolated chemical
anomaly gets extensively stretched when it is carried by magma within
a channel.  Indeed, channels will aggregate magmas vertically, as in
the model here, but will also aggregate laterally by their suction.
The present formulation could be extended to include parameterised
channel flow, but lateral aggregation of diverse melts would require a
two or three-dimensional domain.

Finally, we emphasise that in natural systems, the mantle source and
melt transport are almost certainly heterogeneous. These phenomena
will likely be coupled through lithological heterogeneity of the
source that, by creating productivity heterogeneity, may cause lateral
variability in melt transport rates and structure \citep{lundstrom00,
  kogiso04, weatherley12, katz12}. This potentially creates a complex
interaction between basalt chemistry and its transport through the
mantle. If basalt chemistry is evaluated with this coupled interaction
in mind, then its interpretation in terms of quantitative estimates of
source components becomes more challenging
\citep[e.g.,][]{shorttle14}. However, at a global scale, some
geochemical evidence suggests that major element heterogeneity of the
mantle is relatively inconsequential compared to thermal heterogeneity
\citep[e.g.,][]{gale14}.  Given our limited ability to resolve the
lithologies involved in melting and characterise their melting
behaviours, direct study of the chemical transport associated with a
heterogeneous mantle is not yet tractable.

\subsection{Causes of geochemical variability in basalts}

The present work presents an end-member case that quantifies the
homogenising potential of vertical melt aggregation. Addition of
further complexity in terms of parameterised channel flow would not
serve this purpose and hence has been avoided.  By comparison of our
limited model with observations, we falsify the hypothesis that source
heterogeneity alone (i.e., in the absence of temporal or spatial
heterogeneity of melt transport) can account for variability in melts
delivered from the mantle.

Incremental fractional melts of a homogeneous mantle span a very large
range of concentrations from highly enriched (deepest, incipient
melts) to highly depleted (shallowest melts).  Aggregation with
vertical transport averages this variability.  Channels that transport
deep melts to the surface with limited aggregation of shallower melts
are thus an appealing hypothesis for the observed variability.  Models
of channelised flow \cite[e.g.,][]{aharonov95, spiegelman01}, were
shown by \cite{jull02} and \cite{spiegelman03a} to deliver very large
trace-element variability to the crust.  The present results lend
support to this hypothesis by demonstrating the shortcomings of a
transport model without channelisation.  

Channels emerge because of a positive feedback between vertical flux,
reactive melting, and porosity (permeability) growth.  The magma in
channels is underpressured due to their high permeability and vertical
extent.  This underpressure draws in melts laterally (and also drives
compaction; see \cite{reesjones18b}). Reactive melting persists in
channels until pyroxene has been exhausted from the residue. It
remains unclear whether, in the absence of in situ melting, a lateral
influx of melt is sufficient to maintain open channels at steady state
\citep{liang10b}. Regardless, it is evident that aggregation of melts
occurs even in a channel. The theory presented above should also be
relevant for understanding the consequences of that aggregation.

Moreover, the depth to which channels penetrate remains poorly
constrained \cite[though see][]{jull02}.  It may be impossible for
channels to reach the base of the melting regime, where the
segregation melt flux is small.  If channels penetrate to an
intermediate depth within the melting region, there could be
homogeneous melt transport below that depth.  Trace elements with
sufficiently small $\pco$ would then have a transfer regime that is
entirely deeper than the onset of channels.  For those trace elements,
the model developed here would be useful in predicting how source
heterogeneity is admitted (or attenuated) in deep melts before they
enter channels.

A key factor that complicates these considerations is that the mantle
is heterogeneous in major elements as well as trace elements.  Indeed
source variations of trace and major elements may derive from the same
process and therefore have tight spatial correlation
\citep[e.g.,][]{langmuir1980_rsoc,hirschmann1996_cmp,shorttle11}. Major
element variability affects the fusibility of the mantle, and hence
the distribution of productivity with depth.  Melting of fertile
domains may be fuelled by heat from surrounding, refractory regions
\citep{katz11}.  Melt derived from fertile domains could promote
channelisation \citep[e.g.,][]{lundstrom00, weatherley12, katz12} or
magmatic waves.  \cite{jordan18} has shown that solitary magmatic
waves may be able to trap and transport geochemical signals in
isolation from surrounding melts.  Hence it seems likely that a
comprehensive explanation for geochemical variations in erupted
basalts should account for both source and transport heterogeneity,
and their interaction.  This remains a major challenge.

Clarifying the behaviour of end-member models of geochemical
variability is a useful step toward this goal. Here we have emphasised
the variability of trace-element concentrations, for which there are
many measurements.  A consideration of stable and radiogenic isotopes,
while adding some complexity to the problem, may ultimately be
necessary to disentangle the physical processes involved in melt
extraction from a heterogeneous mantle.  Future models should
incorporate such tracers, and should explore the space of models
that incorporate heterogeneity of both the mantle source and of the
melt transport process.

\appendix{}

\section{Melting column models}
\label{sec:melting-cols}

A melting column is typically defined in the context of a mid-ocean
ridge or mantle plume, where melting occurs as a consequence of
isentropic decompression of the upwelling solid mantle. The column is
a one-dimensional domain, aligned with gravity, in which we solve the
steady, Boussinesq, two-phase equations \citep[e.g.,][]{ribe85a}.
Mass conservation for the liquid and solid phases is expressed as
\begin{subequations}
  \label{eq:meltcol-mass}
  \begin{align}
    \diff{}{z} \left(\phi w\right) &= \Gamma/\rho,\\
    \diff{}{z} \left[(1-\phi)W\right] &= -\Gamma/\rho.
  \end{align}
\end{subequations}
Defining the degree of melting as
\begin{equation}
  \label{eq:define-degree-of-melting}
  F(z) \equiv \frac{\int_{z_0}^{z}\Gamma(z')\,\infd z'}{\rho W_0},
\end{equation}
the mass conservation equations can be integrated to give
\begin{subequations}
  \label{eq:steady-mass-cons}
  \begin{align}
    \label{eq:steady-mass-cons_s}
    (1-\phi) W &= W_0(1-F),\\
    \label{eq:steady-mass-cons_l}
    \phi w &= W_0F.
  \end{align}
\end{subequations}

The momentum conservation equation for the two-phase aggregate is
derived by combining the Darcy-like balance for the liquid phase with
the Stokes-like balance for the solid phase \citep{mckenzie84} to give
\begin{equation}
  \label{eq:meltcol-mom}
  \phi(w-W) = k_\phi(1-\phi)\Delta\rho g/\mu,
\end{equation}
where $\Delta\rho$ is the density difference between solid and liquid,
$g$ is gravitational acceleration, and $\mu$ is the magma viscosity;
the $z$-direction is chosen to be positive upwards (opposite to
gravity). The permeability is given by
\begin{equation}
  \label{eq:permeabilty}
  k_\phi\equiv k_0(\phi/\phi_0)^n,
\end{equation}
where $k_0$ is the permeability when the porosity is equal to the
reference porosity $\phi_0$ and $n=2$ is a constant that we fix
according to empirical and theoretical results for small porosity
\citep[e.g.,][]{vonbargen86, miller14, rudge18}.
Equation~\eqref{eq:meltcol-mom} is derived by making the
zero-compaction-length approximation in which compaction stresses are
neglected relative to Darcy drag \citep{ribe85a, spiegelman93a}.

Combining the integrated mass conservation
equations~\eqref{eq:steady-mass-cons} with~\eqref{eq:meltcol-mom}
gives us the implicit solution written in
equations~\eqref{eq:meltcol-solutions} for $\phi(z),\,w(z),\,W(z)$. A
melting model then determines $\Gamma$ and closes the equations.  In
Appendix~\ref{sec:simplest} we prescribe $\Gamma$ by imposing a
constant isentropic productivity; in
Appendix~\ref{sec:meltcol-volatiles} we develop a melting model for
$\Gamma$ that includes the effect of volatile elements.

\section{A simple column with analytical constraints on admittance}
\label{sec:simplest}

Here we assume that the melting rate is driven by bulk decompression.
In particular, we take $\Gamma = \rho W_0\iprod$, where $\iprod$ is a
constant, uniform, isentropic productivity of upwelling. Then
$F = \iprod (z-z_0) \equiv \iprod Z$.  Here we have defined $Z$ as the
dimensional height above the bottom of the column.

\begin{figure}[ht]
  \centering
  \includegraphics[width=\textwidth]{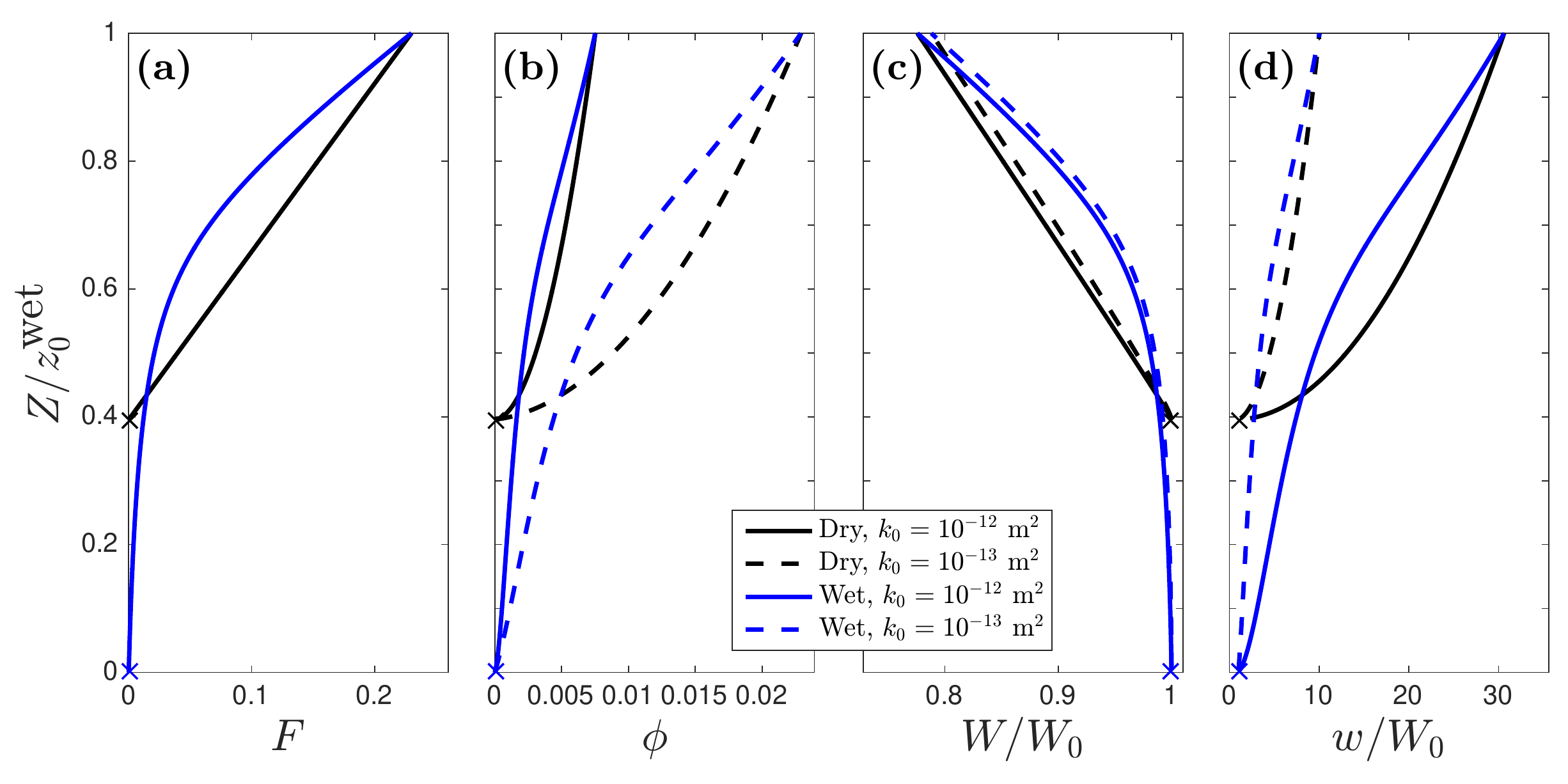}
  \caption{Melting column solutions for the two melting models
    considered here. Black lines show a prescribed, constant
    isentropic productivity (Dry); blue lines show the
    two-component model representing the effect of volatiles
    (Wet). Two values of the reference permeability $k_0$ are
    considered.  \textbf{(a)} Degree of melting; \textbf{(b)}
    Porosity; \textbf{(c)} Solid upwelling rate; \textbf{(d)} Liquid
    upwelling rate. Parameters for all curves include: permeability
    exponent $n = 2$; viscosity $\mu = 10$~Pa~s; density with the
    Boussinesq assumption $\rho = 3000$~kg/m$^3$; density difference
    between the solid and liquid phase $\Delta\rho = 500$~kg/m$^3$;
    $\Fmax=0.23$ and $W_0 = 4$~cm/yr. The volatile here is water with
    a partition coefficient $\pco_w = 0.01$; source volatile
    concentration $100$~ppm; heat capacity $c_P=1200$~J/K;
    Clausius-Clapeyron slope $\clapeyron=5.54\times10^{-6}$~Pa/K;
    specific latent heat $L=10^{6}$~J/kg; coefficient of thermal
    expansion $\alpha=3 \times 10^{-5}$~K$^{-1}$; volatile solidus
    depression $M = -4$~K/ppm; reference mantle temperature
    $\Tsol_0=1600$~K.}
  \label{fig:dry_wet_melting_col}
\end{figure}

An explicit, analytical solution to
equations~\eqref{eq:meltcol-solutions} can be obtained for $n=2$ or
$3$. The former is more appropriate at very small porosity
\citep{rudge18}. In Figure~\ref{fig:dry_wet_melting_col}, we plot the
$n=2$ solution to the system of
equations~\eqref{eq:meltcol-solutions}.

With this column model and neglecting exchange reactions ($\rxn=0$),
equations~\eqref{eq:mean} for the mean concentration become
\begin{subequations}
  \label{eq:simple_mean_gov}
  \begin{align}
    \label{eq:simple_mean_gov_s}
    \diff{\csm}{F} &= -\left(1/\pco-1\right)\frac{\csm}{1-F},\\
    \label{eq:simple_mean_gov_l}
    \diff{\clm}{F} &= -\frac{\clm - \csm/\pco}{F}.    
  \end{align}
\end{subequations}
Application of the time-independent part of the boundary
condition~\eqref{eq:boundary_condition} yields the solution
\begin{subequations}
  \begin{align}
    \label{eq:simple_solid_mean_soln}
    \csm(F) &= \csm_0\left(1-F\right)^{1/\pco-1},\\
    \label{eq:simple_liquid_mean_soln}
    \clm(F) &= \csm_0\frac{1-\left(1-F\right)^{1/\pco}}{F}.
  \end{align}
\end{subequations}
These are the equations of aggregated fractional melting
\citep{shaw06}.

The equations for the fluctuation amplitudes follow from
\eqref{eq:fluctuations} using mass-conservation
equations~\eqref{eq:steady-mass-cons},
\begin{subequations}
  \begin{align}
    \label{eq:simple_fluct_gov_s}
    \diff{\csf}{Z} &= -\frac{\iprod\csf}{1-\iprod
                     Z}\left(\frac{1}{\pco} - 1\right) -
                     i\frac{\freq\csf}{W(Z)},\\
    \label{eq:simple_fluct_gov_l}
    \diff{\clf}{Z} &= -\frac{\clf - \csf/\pco}{Z} - 
                     i\frac{\freq\clf}{w(Z)}.
  \end{align}
\end{subequations}
Using the fluctuating part of the boundary
condition~\eqref{eq:boundary_condition} and $F=\iprod Z$ gives the
solution
\begin{subequations}
  \label{eq:simple_fluct_soln}
  \begin{align}
    \label{eq:simple_fluct_soln_s}
    \csf(F) &= \csf_0\e^{-i\freq t^s(F)}\left(1-F\right)^{1/\pco - 1},\\
    \label{eq:simple_fluct_soln_l}
    \clf(F) &= \csf_0F^{-1}\e^{-i\freq t^\ell(F)}
              \int_0^F\e^{-i\freq\Delta t(F')}\frac{1}{D}\left(1-F'\right)^
              {1/\pco - 1}\infd F'.
  \end{align}
\end{subequations}
Here we have introduced 
\begin{equation}
  \label{eq:transport_times}
  t^s(F) \equiv \int_0^F \frac{1}{W(F')}\frac{\infd F'}{\iprod},\qquad
  t^\ell(F) \equiv \int_0^F \frac{1}{w(F')}\frac{\infd F'}{\iprod},
\end{equation}
as the time for a parcel of solid or liquid, respectively, to move
from the bottom of the melting column ($F=0$) to the height where the
degree of melting is $F$.  We also defined
$\Delta t \equiv t^s - t^\ell$. These definitions allow us to avoid
specifying upwelling rates, for the moment.

Using the solution~\eqref{eq:simple_fluct_soln}, we can evaluate the
admittances defined in equation~\eqref{eq:def_admittance}. For the
solid phase,
\begin{equation}
  \label{eq:simple_admittance_s}
    \admits(F) = \left(1 - F\right)^{1/\pco - 1}.
\end{equation}
From this expression and the
solution~\eqref{eq:simple_solid_mean_soln}, it is evident that at any
height in the column, $\admits$ is equal to $\csm(Z) / \csm_0$. This
means that for the solid phase, the decay of concentration
fluctuations with height in the melting column is identical to the
decay of the mean concentration.

For the liquid phase,
\begin{equation}
  \label{eq:simple_admittance_l_integral}
  \admitl =
  \frac{\left\vert\clf(F)\right\vert}{\left\vert\csf_0\right\vert}
  = \frac{1}{F}\left\vert\int_0^{F}\e^{-i\freq\Delta t(F')}
    \frac{1}{\pco}\left(1-F'\right)^{1/\pco - 1}\infd F'\right\vert.
\end{equation}
This equation cannot be evaluated analytically without
approximations. It can, however, be bounded according to
\begin{equation}
  \label{eq:simple_admittance_l}
    \admitl \leq \frac{1}{\pco Z}\int_0^Z\left\vert\left(1-\iprod
        Z\right)^{\left[1/\pco - 1 + i\freq/(\iprod W_0)\right]}
      e^{i\freq \int_0^Z (1/w) \infd Z}\right\vert\infd Z 
    = \frac{1-\left(1-\iprod Z\right)^{1/\pco}}{\iprod Z},
\end{equation}
where we have used $F=\iprod Z$ and the integral inequality
$\left\vert \int f\,\infd z\right \vert \leq \int\left \vert f\right
\vert\left \vert\infd z\right \vert$. Comparing
\eqref{eq:simple_admittance_l} with the
solution~\eqref{eq:simple_liquid_mean_soln} for the mean liquid
concentration,
\begin{equation}
  \label{eq:simple_admittance_l_versus_mean}
  \admitl(F) \leq \clm(F)/\csm_0.
\end{equation}
Therefore we conclude that for the liquid phase, the decay of
concentration fluctuations with height in the melting column is at
least as rapid as the decay of the mean liquid
concentration. 

We make further progress by introducing assumptions that simplify the
integrand of~\eqref{eq:simple_admittance_l_integral}. We first
consider the quantity $\Delta t$, which represents the time difference
for transport of the solid and liquid phases between $Z=0$ and
$Z=F/\iprod$. It is expanded as
\begin{equation}
  \label{eq:delta-t-model}
  \Delta t(F) = \int_0^F\left(\frac{1}{W(F')} -
    \frac{1}{w(F')}\right)\frac{\infd F'}{\iprod} \equiv \dura(F),
\end{equation}
where $\dura(F)$ is an unknown, nonlinear function. When
$k_0\to\infty$, the permeability is infinite and $1/W \gg 1/w$ at all
$F>0$.  In this limiting case $t^\ell\sim 0$ and we find,
using~\eqref{eq:steady-mass-cons_s}, that
$\dura(F) \sim -\ln(1-F)/(W_0\iprod)$.  This can be further simplified
when $F \ll 1$ to give
\begin{equation}
  \label{eq:asymptotic-linear-coef}
  \dura(F) \sim \frac{F}{W_0\iprod} \equiv \dura_0 F.
\end{equation}
This result simply means that the travel time difference at any height
$Z$ is approximated by the travel time of the solid at the background
upwelling speed: $\Delta t \sim (z-z_0)/W_0$.

We also introduce the approximation
$\left(1-F\right)^{1/\pco - 1} \approx \e^{-F/\pco}$, which requires
the additional assumption that $\pco$ is much smaller than unity. For
the purposes of this manuscript, it is adequate that
$\pco \lesssim 0.1$ for $F\lesssim 0.2$.

Using the these approximations to re-write
equation~\eqref{eq:simple_admittance_l_integral} gives
\begin{equation}
  \label{eq:admittance-simplified-integral}
  \admitl \sim \frac{1}{F}\left\vert\int_0^{F}\frac{\e^{-(i\freq\dura_0
    + 1/\pco)F'}}{\pco}\infd F'\right\vert.
\end{equation}
This integral can be evaluated to give
\begin{equation}
  \label{eq:admittance-asymptotic-model}
  \admitl \sim \frac{1}{F}\frac{\sqrt{\left(1-\e^{-F/\pco}\right)^2 
      + 4\e^{-F/\pco}\sin^2\left(F\dura_0\freq/2\right)}}
  {\sqrt{1 + (\pco\dura_0\freq)^2}}.
\end{equation}
Recall that this asymptotic result is strictly valid for $\pco,F\ll 1$
and $k_0\to\infty$. For highly incompatible elements with
$\pco\ll F$, we can simplify further and obtain
\begin{equation}
  \label{eq:admittance-asymptotic-model-highly-incompat}
  \admitl \sim \frac{1}{F\sqrt{1 +
      (\pco\dura_0\freq)^2}}\qquad\left(\text{for~}\pco\ll F\right).
\end{equation}
The simple approximations \eqref{eq:admittance-asymptotic-model} and
\eqref{eq:admittance-asymptotic-model-highly-incompat} capture the
structure of the admittance well when $\rr=0$ and the melt
productivity is constant. They are plotted in
Figures~\ref{fig:trace_columnview} and \ref{fig:admittance_dry} and
discussed in section~\ref{sec:admittance-dry} of the main text.

\section{A melting column with volatiles}
\label{sec:meltcol-volatiles}

To incorporate the effect of volatile elements on the steady-state
porosity and velocity profiles in the column, we append a simple
thermochemical model to equations~\eqref{eq:meltcol-mass} and
\eqref{eq:meltcol-mom}, following the approach of
\cite{reesjones18a}. This uses a steady-state conservation of energy,
written in terms of temperature $T$ as
\begin{equation}
  \label{eq:meltcol-energy}
  \rho c_PW_0\diff{T}{z} = -(L\Gamma + \rho\alpha W_0 g T),
\end{equation}
where $c_P$ is specific heat capacity, $L$ is the latent heat of
fusion (J/kg) and $\alpha$ is the coefficient of thermal
expansion. This equation states that the advection of sensible heat by
bulk upwelling of rock and magma is balanced by conversion to latent
heat through melting, and conversion to work through volume expansion.

We assume that the mantle is composed of two components, ``rock'' and
``volatiles.'' Volatile concentration is denoted by a $\vcon$
(capitalised) to distinguish it from a trace-element
concentration. The solidus is the relationship between temperature,
pressure and the volatile concentration of the solid when both solid
and melt are present. We assume a simple form in which this
relationship is linearised about the conditions at the bottom of the
column,
\begin{equation}
  \label{eq:solidus}
  T = \Tsol_0 - \clapeyron^{-1}\rho g (z-z_0) - M (\vcon^s-\vcon^s_0),
\end{equation}
where $\clapeyron$ is the Clausius-Clapeyron slope and $M$ is the
slope of the solidus with volatile concentration; both are taken as
constants. We assume a lithostatic pressure gradient.  The liquidus
curve is defined by the assumption of a constant ratio,
$\pco_w \equiv \left[\vcon^s/\vcon^\ell\right]$, between the volatile
concentration in the solid and the liquid.  Hence the equilibrium
compositional difference between phases is
$\Delta\vcon = \vcon^s(1-1/\pco_w)$.  Using the lever rule referenced
to the initial concentration, we can define the degree of melting of
the solid phase as
$F \equiv \left(\vcon^s - \vcon^s_0\right)/\Delta\vcon$.  Then,
combining this with equation~\eqref{eq:solidus} and the partitioning
relation for $\Delta\vcon$, we can express $F$ as a function of
temperature,
\begin{equation}
  \label{eq:F_of_T}
  F = \frac{T-\Tsol_0 + \clapeyron^{-1}\rho g z}{(1-1/\pco_w)(T-\Tsol_0 +
    \clapeyron^{-1}\rho g z + M\vcon^s_0)}.
\end{equation}

Using \eqref{eq:F_of_T}, the melting rate $\Gamma$ can be written as 
\begin{equation}
  \label{eq:meltingrate_F}
  \Gamma = \rho W_0\diff{F}{z} = \rho W_0\pdiff{F}{T}\left(\diff{T}{z}
    + \clapeyron^{-1}\rho g\right).
\end{equation}
Substituting this into the conservation of energy equation
\eqref{eq:meltcol-energy} gives
\begin{equation}
  \label{eq:2}
  \left(1 + \frac{L}{c_P}\pdiff{F}{T}\right)\diff{T}{z} + \frac{\alpha
    g}{c_P}T + \frac{\rho g \clapeyron^{-1} L}{c_P}\pdiff{F}{T} = 0.
\end{equation}
Since $\partial F/\partial T$ is a function $T$, this equation is
nonlinear; we integrate it numerically to find $T(z)$, which is then
used in \eqref{eq:F_of_T} to find $F(z)$.  Mass and momentum
conservation are then used to obtain $\phi,w,W$.  Two example
solutions with different values of $k_0$ are shown in
Figure~\ref{fig:dry_wet_melting_col}.

Properties of the admittance of trace-element variations are not
available by analytical methods for this melting column.  We obtain
results by numerical methods in section~\ref{sec:admittance-wet} of
the main text.

\section{Non-compacting boundary layer}
\label{sec:noncom-boundary}

All the above melting column solutions are based on the
zero-compaction-length (ZCL) approximation, which neglects gradients
in the compaction pressure.  These gradients are important only in a
narrow boundary layer at the bottom of the melting column
\citep[e.g,][]{ribe85a,sramek07}. However, it is precisely at the
bottom of the melting column (in the transfer regime) where
attenuation of heterogeneity occurs.  Therefore, it is important to
consider whether the ZCL approximation makes a qualitative difference
to the results and conclusions of this study.

Within the narrow boundary layer near the onset of melting, melt
buoyancy is balanced by a gradient in the compaction pressure; there
is little compaction and hence $w\sim W_0$ and $\phi\sim F$ (in the
rest of the column, buoyancy is balanced by Darcy drag). Therefore,
the ZCL approximation predicts liquid segregation that is too rapid,
compared to the full solution, in the boundary layer.  We have shown
that the key factor controlling admittance is the accumulated phase
difference in the transfer regime. Hence, we need to evaluate the
importance of the non-compacting boundary layer (NCBL) within the
transfer regime.

When the height of transfer regime is much larger than the height of
the non-compacting boundary layer, e.g., for a mildly incompatible
trace-element, the inaccuracy of the ZCL assumption is clearly
negligible. For small enough $\pco$, the height of transfer regime
will become comparable height to the NCBL. In this case, the phase
difference would be reduced within the transfer regime,
which would diminish the attenuation.  But if the wavelength of
heterogeneity is much larger than both the NCBL and the transfer
regime height, then attenuation is minimal anyway. These two cases
cover all combinations of $\pco$ and $\lambda_0$ considered here and
hence the ZCL approximation doesn't qualitatively affect our results.

\begin{figure}[ht]
  \centering
  \includegraphics[width=\textwidth]{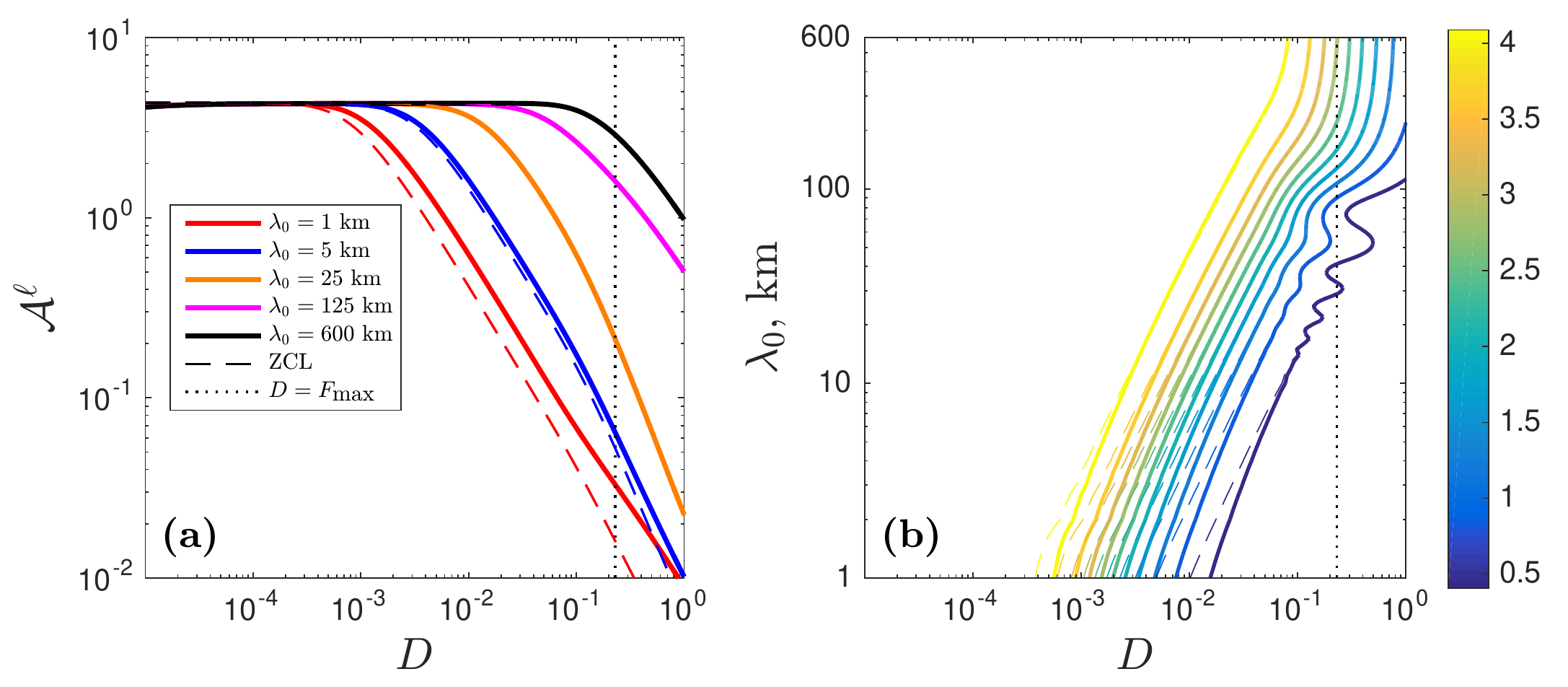}
  \caption{Admittance $\admitl$ calculated for a dry melting column
    with a numerical solution that considers both the compaction
    stress and the Darcy drag. All other parameters are identical to
    Fig.~\ref{fig:admittance_dry}. \textbf{(a)} $\admitl$ as a
    function of partition coefficient $\pco$ for various wavelengths
    of heterogeneity, as in legend. \textbf{(b)} Contours of constant
    $\admitl$ as a function of $\pco$ and input heterogeneity
    wavelength $\lambda_0$.  Other parameters are the same as in
    Figure~\ref{fig:trace_columnview}}
  \label{fig:admittance_noncom_boundary}
\end{figure}

To demonstrate that the quantitative effect is small, we have computed
numerical solutions of melt segregation that don't neglect gradients
in compaction stress (in the dry case only).  These results are used
in the trace element model to compute the admittance. Figure
\ref{fig:admittance_noncom_boundary} shows the change in admittance
from relaxing the ZCL assumption, for all other parameters held
constant.

\paragraph{Acknowledgements} No data were harmed (or created) in the
making of this study. The authors thank C.~Ballentine for helpful
suggestions and reviewers A.~Stracke, Y.~Liang and J.~Jordan for
insightful comments. The research leading to these results has
received funding from the European Research Council under the European
Union’s Seventh Framework Programme (FP7/2007–2013)/ERC grant
agreement number 279925, as well as the NERC Volatiles Consortium
under grant NE/M000427/1. J.F.R.~thanks the Leverhulme Trust for
support.

\bibliographystyle{abbrvnat}
\bibliography{manuscript}

\end{document}